\documentclass[12pt,article,onecolumn]{IEEEtran}

\usepackage{float}
\usepackage{graphicx}
\usepackage{amssymb}
\usepackage{epsfig}
\usepackage{amsmath}
\usepackage{amsthm}
\usepackage{amsfonts}
\usepackage{setspace}
\usepackage{url}
\usepackage{psfrag}
\usepackage{graphicx}
\usepackage{dsfont}
\usepackage{cite}

\usepackage{units}

\newtheorem{thm}{Theorem}
\newtheorem{cor}{Corollary}
\newtheorem{lem}{Lemma}

\newtheorem{defn}{Definition}

\newcommand{\E}{\textnormal{E} }

\begin{document}

\title{Dispersion of Infinite Constellations in Fast Fading Channels}

\author{Shlomi Vituri, and Meir Feder,~\IEEEmembership{Fellow,~IEEE}
\thanks{The authors are with the Department of Electrical Engineering --
Systems, Tel-Aviv University, Ramat-Aviv 69978, Israel (e-mails:
viturish@post.tau.ac.il,meir@eng.tau.ac.il).}}

\maketitle

\begin{abstract}
In this work we extend the setting of communication without power constraint, proposed by Poltyrev, to fast fading channels with channel state information (CSI) at the receiver.
The optimal codewords density, or actually the optimal normalized log density (NLD), is considered.
Poltyrev's capacity for this channel is the highest achievable NLD, at possibly large block length, that guarantees a vanishing error probability.
For a given finite block length $n$ and a fixed error probability $\epsilon$, there is a gap between the highest achievable NLD and Poltyrev's capacity.
As in other channels, this gap asymptotically vanishes as the square root of the channel dispersion $V$ over $n$, multiplied by the inverse Q-function of the allowed error probability. This dispersion, derived in the paper, equals the dispersion of the power constrained fast fading channel at the high SNR regime.
Connections to the error exponent of the peak power constrained fading channel are also discussed.
\end{abstract}
\IEEEpeerreviewmaketitle
\section{Introduction}
Wireless communication channels are traditionally modeled as fading channels, where the transmitted signal is multiplied by a fading process and observed with additive white Gaussian noise (AWGN). In a fast fading channel the fading process is composed of fading coefficients, modeled as independent and identically distributed (i.i.d.) random variables. This is a reasonable model for many practical wireless communication systems, such as systems that use a (pseudo) random interleaver between the transmitted digital symbols (e.g. BICM techniques) over, e.g., a frequency selective wireless channel. Here we will assume that a perfect knowledge of the channel state information (the fading coefficients) is available at the receiver.

Classical coding problems over the fading channels often include a peak or an average power restriction of the transmitted signal. Without power constraint the capacity of the channel is not limited, since we can choose an infinite number of codewords to be arbitrarily far apart from each other, and hence get an arbitrarily small error probability and infinite rate. Nevertheless, coded modulation methods ignore the power constraint by designing infinite constellations (IC), and then taking only a subset of codewords which are included in some ``shaping region'' to get a finite constellation (FC) that holds the power constraint. Hence, IC is a very convenient framework for designing codes.

Poltyrev studied in \cite{Poltyrev} the IC performance over the AWGN without power constraint. He defined the density (the average number of codewords per unit volume) and the normalized log density (NLD) of the IC, in analogy to the number of codewords and the communication rate in the power constraint model, respectively. He showed that the highest achievable NLD over the unconstrained AWGN channel, with arbitrarily small error probability, is limited by a maximal NLD, sometimes termed the 'Poltyrev's capacity'. He also derived an exact term for the maximal NLD and error exponent bounds using random coding and sphere packing techniques, for any NLD below the capacity.

In classical channel coding problems, the capacity gives the maximal achievable communication rate when arbitrarily small error probability is required (and arbitrary large codeword length $n$ is permitted). The error exponent provides the exponential rate of convergence (with $n$) in which the error probability goes to zero, for any fixed rate below the capacity. Another interesting question is: for a fixed error probability $\epsilon$ and a fixed codeword length $n$, what is the maximal achievable rate, denoted by $R^*(n,\epsilon)$. Although this question is still unsolved for any finite $n$, the recently revisited dispersion analysis \cite{Polyanskiy} gives the rate of convergence of $R^*(n,\epsilon)$ to the capacity. According to the dispersion analysis, for any fixed $\epsilon$ and finite $n$ the following holds:
\begin{equation}
\label{eq_dispersion_analysis}
R^*(n,\epsilon) = C - \sqrt{\frac{V}{n}}Q^{-1}(\epsilon) + O\left(\frac{\ln(n)}{n}\right),
\end{equation}
where $Q$ is the standard complementary Gaussian CDF, $C$ is the channel capacity and $V$ is the channel dispersion. The channel dispersion is given by the variance of the information density $i(x;y) \triangleq \ln\left(\frac{P(x,y)}{P(x)P(y)}\right)$ for a capacity achieving input distribution. Polyanskiy et al. showed in \cite{Polyanskiy} that \eqref{eq_dispersion_analysis} holds for discrete memoryless channels (DMCs) and for AWGN channel. In \cite{Polyanskiy_fading} the result was extended to stationary fading channels.

In \cite{Ingber} Ingber et al. showed that in AWGN channel without power constraint and with noise variance $\sigma^2$, the analogy of \eqref{eq_dispersion_analysis} for IC is given by:
\begin{equation}
\label{eq_dispersion_analysis_IC}
\delta^*(n,\epsilon) = \delta^* - \sqrt{\frac{V}{n}}Q^{-1}(\epsilon) + O\left(\frac{\ln(n)}{n}\right),
\end{equation}
where $\delta^*(n,\epsilon)$ is the optimal NLD for fixed $\epsilon$ and finite $n$, and $\delta^* \triangleq \frac{1}{2}\ln\left(\frac{1}{2\pi e\sigma^2}\right)$ is Poltyrev's capacity. For AWGN, the channel dispersion is given by $V = \frac{1}{2}$, which is equal to the limit of the channel dispersion of the power constrained AWGN, when the SNR tends to infinity.

In this paper we extend Poltyrev's setting to the case of a fast fading channel with AWGN and without power constraint. The main result of this paper is that an analogous expression to \eqref{eq_dispersion_analysis_IC} holds for fast fading channels. Moreover, the dispersion of unconstrained fast fading channel, derived later in the paper, equals the limit of the dispersion of the fast fading channel with power constraint, derived in \cite{Polyanskiy_fading}, when the SNR tends to infinity.

In the achievability part of the proof, we will use the \emph{Dependence Testing Bound} that was used in \cite{Polyanskiy} to prove the achievability part of \eqref{eq_dispersion_analysis} for DMCs. This bound is based on random coding and on a suboptimal decoder. The suboptimal decoder is based on information density threshold crossing. Here, we will use this bound for bounding the average error probability over the ensemble of codes with codewords that are uniformly distributed on an $n$-dimensional cube with length $a$. By letting $a$ tend to infinity, we will prove the existence of an IC with NLD that is lower bounded by the right hand side (RHS) of \eqref{eq_dispersion_analysis_IC}. In the converse part of the proof, we will use the sphere packing bound for the average error probability and its asymptotical distribution for large $n$, in fast fading channels.

The paper is organized as follows. In section \ref{sec_basic_definitions} notations and basic definitions for the fading channel model and for IC's are given. In section \ref{sec_relation_to_the_power_constrained_model} connections to the power constrained channel model are discussed. In section \ref{sec_main_result} our main result is presented and proved. In section \ref{sec_extension_to_the_complex_model} we briefly extend our main result to the complex channel model. Finally, we summarize the paper in section \ref{sec_summary}.
\section{Basic Definitions}
\label{sec_basic_definitions}
\subsection{Notation}
Vectors are denoted by bold-face lower case letters, e.g. $\mathbf{x}$ and $\mathbf{y}$. Matrices are denoted by bold-face capital letters, e.g. $\mathbf{H}$. Components of random vector $\mathbf{x}$ are denoted by capital letters, $X_1, X_2, \dots, X_n$. In the same manner, components of a random matrix $\mathbf{H}$ are denoted by $\left\{H_{ij}\right\}$. Instances of random variables (RVs) are denoted by lower case letters, e.g. $x, ~y$ and $h$.
\subsection{Channel Model}
The fast fading channel model is given by
\begin{equation}
\label{eq_channel_model}
Y_i = H_i\cdot X_i + Z_i, ~i = 1,2,\dots
\end{equation}
where,
\begin{itemize}
  \item $\{X_i\}$ is a series of channel inputs,
  \item $\{H_i\}$ is a series of i.i.d. fading coefficients satisfying $E\{H_i^2\}=1$,
  \item $\{Z_i\}$ is a series of i.i.d. normal random variables, such that $Z_i\sim N(0,\sigma^2)$,
  \item $\{Y_i\}$ is a series of channel outputs.
\end{itemize}
The RVs $\{X_i\}, ~\{H_i\}$ and $\{Z_i\}$ are independent of each other. In vector notation (for finite $n$) the channel model is given by:
\begin{equation}
\label{eq_vectoric_channel_model}
\mathbf{y} = \mathbf{H} \cdot \mathbf{x} + \mathbf{z},
\end{equation}
where $\mathbf{H} \triangleq \text{diag}\left\{H_1, H_2, \dots, H_n\right\}$. We assume a perfect CSI available at the receiver, and hence the receiver's channel output is the couple $\left(\mathbf{y},\mathbf{H}\right)$.

Without loss of generality, since we have a perfect CSI at the receiver, we can assume that the fading coefficients are nonnegative. Moreover, we restrict the fading distribution to probability density functions (PDF) with zero probability to equal zero. We will denote such a fading distribution by \emph{regular fading distribution}, which is defined formally below.
\begin{defn}
(Regular fading distribution): A fading PDF $f\left(h\right)$ is called regular fading distribution if there exists some positive constant $\alpha$, s.t. $f\left(h\right) \propto \frac{1}{h^{1-\alpha}}$ for small enough $h > 0$.
\end{defn}
A popular statistical model for the fading channel is the Nakagami-$m$ distribution. It is easy to verify that this distribution is a \emph{regular fading distribution} for all $m \geq \frac{1}{2}$.

In this paper, we investigate the dispersion of regular fading channels in finite dimensional IC with available CSI at the receiver.
\subsection{Infinite Constellations}
An infinite constellation of dimension $n$ is any countable set of points $S = \left\{s_1,s_2,\dots\right\}$ in $\mathbb{R}^n$.

Let $\text{Cb}(a)$ denote an $n$ dimensional hypercube in $\mathbb{R}^n$:
\begin{equation}
\text{Cb}(a) \triangleq \left\{\mathbf{x}\in \mathbb{R}^n ~ s.t. ~ \forall_i \left|x_i\right| < \frac{a}{2} \right\}.
\end{equation}

We denote by $M\left(S,a\right) = \left|S\bigcap\text{Cb}(a)\right|$ the number of points in the intersection of $\text{Cb}(a)$ and $S$.

The density of points per unit volume of $S$ is denoted by $\gamma$ and defined by
\begin{equation}
\gamma \triangleq \limsup_{a \rightarrow \infty} \frac{M\left(S,a\right)}{a^n}.
\end{equation}

The normalized log density of $S$ is denoted by $\delta$ and defined by
\begin{equation}
\delta \triangleq \frac{1}{n}\ln\left(\gamma\right).
\end{equation}

In the receiver, given the channel state information, the receiver's IC, denoted by $S_\mathbf{H}$, is defined by
\begin{equation}
S_\mathbf{H} \triangleq \left\{s_{\text{rc}}: s_{\text{rc}} = \mathbf{H}\cdot s, s \in S\right\}.
\end{equation}

We also define the set ${\mathbf{H}}\cdot\text{Cb}(a)$ as the multiplication of each point in $\text{Cb}(a)$ with the matrix $\mathbf{H}$.

The density of $S_\mathbf{H}$ is defined by
\begin{align}
\gamma_{\text{rc}}\left(\mathbf{H}\right) &\triangleq \limsup_{a \rightarrow \infty} \frac{M\left(S_{\mathbf{H}},a\right)}{\text{Vol}\left(\mathbf{H}\cdot\text{Cb}(a)\right)}
\\ &= \limsup_{a \rightarrow \infty} \frac{M\left(S,a\right)}{\det{\left(\mathbf{H}\right)} \cdot a^n}
\\ &= \frac{\gamma}{\det{\left(\mathbf{H}\right)}}
\end{align}
where $M\left(S_{\mathbf{H}},a\right) \triangleq \left|S_{\mathbf{H}}\bigcap{\mathbf{H}}\cdot\text{Cb}(a)\right|$.

For $s_{\text{rc}} \in S_{\mathbf{H}}$, let $P_e\left(s_{\text{rc}}|\mathbf{H}\right)$ denote the error probability when $s$, such that $s_{\text{rc}} = \mathbf{H} \cdot s$, was transmitted and the CSI at the receiver is $\mathbf{H}$. Then, using maximum likelihood (ML) decoding the error probability is given by
\begin{equation}
P_e\left(s_{\text{rc}}|\mathbf{H}\right) = Pr\left\{s_{\text{rc}} + \mathbf{z} \notin W\left(s_{\text{rc}}\right) | \mathbf{H} \right\}
\end{equation}
where $W\left(s_{\text{rc}}\right)$ is the Voronoi cell of $s_{\text{rc}}$, i.e. the convex polytope of the points that are closer to $s_{\text{rc}}$ than to any other point $s^{'}_{\text{rc}} \in S_{\mathbf{H}}$.
\begin{defn}
(Conditional expectation over a faded hypercube): For any function $f:S_{\mathbf{H}}\rightarrow\mathbb{R}$, the conditional expectation of $f(s_{\emph{\text{rc}}})$ given $\mathbf{H}$, where $s_{\emph{\text{rc}}}$ is drawn uniformly from the code points that reside in the faded hypercube ${\mathbf{H}}\cdot\emph{\text{Cb}}(a)$, will be denoted and defined by
\begin{equation}
E_{S,a|\mathbf{H}}\left\{ f(s_{\emph{\text{rc}}}) \right\} \triangleq \frac{1}{M\left(S_{\mathbf{H}},a\right)} \sum_{s_{\emph{\text{rc}}} \in S_{\mathbf{H}}\bigcap{\mathbf{H}}\cdot\emph{\text{Cb}}(a)}f(s_{\emph{\text{rc}}}).
\end{equation}
\end{defn}
The average error probability using ML decoding and equiprobable messages transmission is given by
\begin{align}
P_e\left(S\right) = E\left\{ P_e\left(S_{\mathbf{H}}\right) \right\} &\triangleq E\left\{ \limsup_{a \rightarrow \infty} \frac{1}{M\left(S_{\mathbf{H}},a\right)} \sum_{s_{\text{rc}} \in S_{\mathbf{H}}\bigcap{\mathbf{H}}\cdot\text{Cb}(a)} P_e\left(s_{\text{rc}}|\mathbf{H}\right) \right\}
\\                &\triangleq E\left\{ \limsup_{a \rightarrow \infty} E_{S,a|\mathbf{H}}\left\{ P_e\left(s_{\text{rc}}|\mathbf{H}\right) \right\} \right\}.
\end{align}
\section{Relation to the Power Constrained Model}
\label{sec_relation_to_the_power_constrained_model}
The error exponent at rates near the capacity can be approximated by a parabola of the form
\begin{equation}
\label{eq_error_exponent_near_capacity}
\E\left(R\right) \approx \frac{\left(C-R\right)^2}{2V},
\end{equation}
where $V$ is the channel dispersion. This fact was already known to Shannon (see Fig. 18 in \cite{Polyanskiy}). By taking uniform input distribution in Gallager's random coding error exponent \cite{Gallager}, precisely $X \sim U\left(-\frac{a}{2},\frac{a}{2}\right)$, over the power constrained fast fading channel with available CSI at the receiver, it can be shown that \eqref{eq_error_exponent_near_capacity} holds with $C = E\left\{\frac{1}{2}\ln\left( \frac{a^2H^2}{2\pi e\sigma^2} \right)\right\}$ and $V = \frac{1}{2} + Var\left(\frac{1}{2}\ln\left(H^2\right)\right)$, when $a/\sigma$ tends to infinity (the high SNR regime). Since the unconstrained setting can be thought of as the limit of the power constrained setting when the SNR tends to infinity, this result hints that $\delta^* = E\left\{ \frac{1}{2}\ln\left(\frac{H^2}{2\pi e\sigma^2}\right) \right\}$ and $V = \frac{1}{2} + Var\left(\frac{1}{2}\ln\left(\frac{H^2}{2\pi e\sigma^2}\right)\right)$, in that setting.

In \cite{Polyanskiy_fading} Polyanskiy et al. studied the dispersion of the general case of power constrained stationary fading channels. He showed that the dispersion is affected by the fading dynamics, this is in contrary to the channel capacity, which is independent of this dynamics \cite{Fading_channels}. Moreover, in some fading processes, such as Gauss-AR processes, this dispersion is increased relative to fast fading channel with the same marginal fading distribution. This fact can motivate the useage of random interleaver in practical systems with finite block-length, in order to get effectively a fast fading channel (with smaller channel dispersion). In case of fast fading channels with power constraint $P$, and AWGN variance $\sigma^2$, this dispersion (in $\text{nats}^2$ per channel use) is given by
\begin{equation}
\label{eq_dispesion_fading_power_constrained}
V = Var\left(\frac{1}{2}\ln\left(1+SNR\cdot H^2\right)\right) + \frac{1}{2}\left(1 - E^2\left\{\frac{1}{1+SNR\cdot H^2}\right\}\right),
\end{equation}
where $SNR \triangleq P/\sigma^2$. Another indication to the channel dispersion value in the unconstrained case is given by taking the limit of \eqref{eq_dispesion_fading_power_constrained} when the SNR tends to infinity. In the high SNR regime \eqref{eq_dispesion_fading_power_constrained} can be approximated by
\begin{align}
V &\approx \frac{1}{2} + Var\left(\frac{1}{2}\ln\left(SNR\cdot H^2\right)\right)
\\&= \frac{1}{2} + Var\left(\frac{1}{2}\ln\left(H^2\right)\right),
\end{align}
which coincides with the previous hint to the channel dispersion value in the unconstrained setting. The case of stationary fading channels with memory is a subject for further research.

It should be noted that while the dispersion analysis accuracy of power constrained fading channels in \cite{Polyanskiy_fading} is $o\left(\frac{1}{\sqrt{n}}\right)$, in our analysis the accuracy is slightly better, $O\left(\frac{\ln(n)}{n}\right)$. This faster convergence might be due to the fact that in \cite{Polyanskiy_fading} a more general fading model was analyzed.

\begin{figure}[htp]
\center{\includegraphics[width=1\columnwidth]{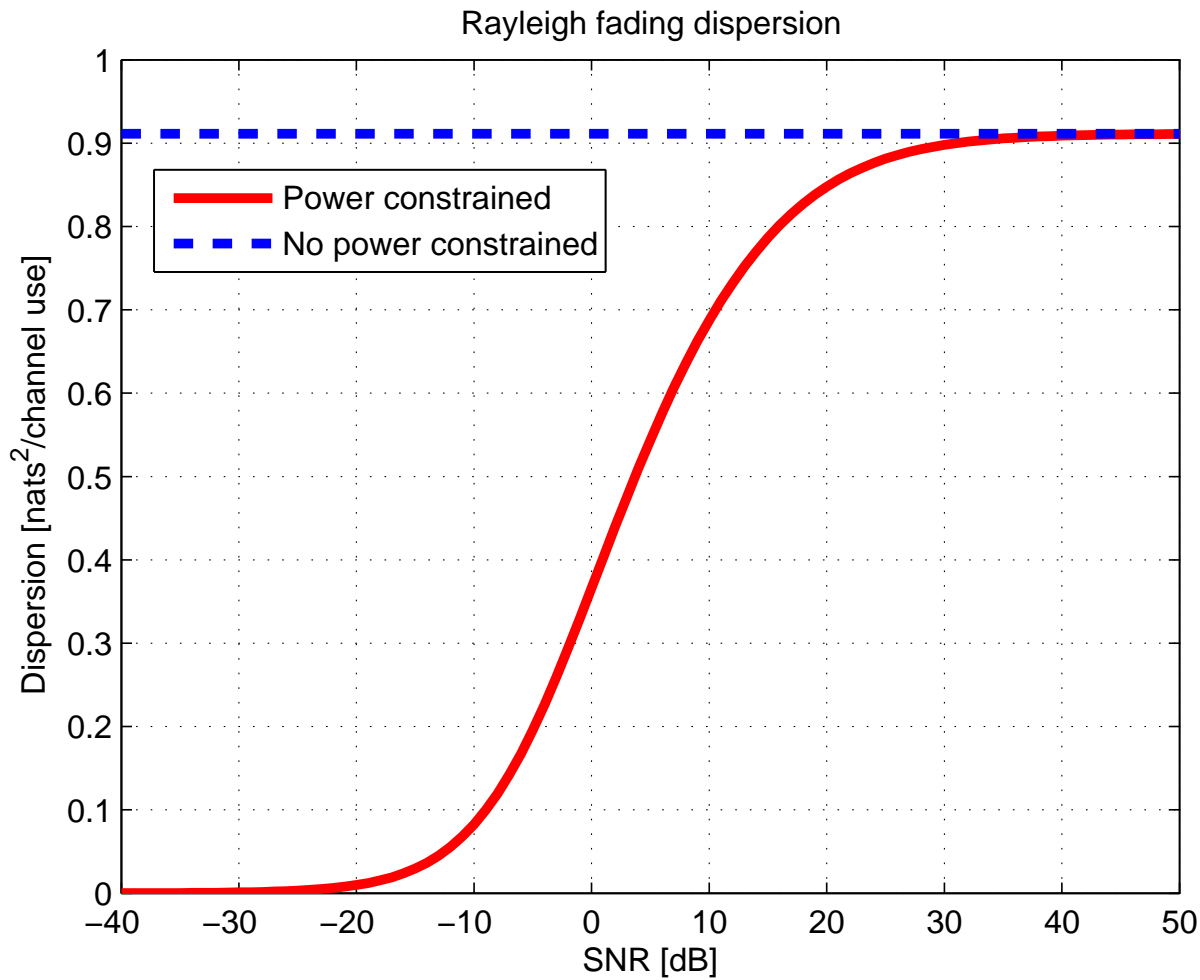}}
\caption{\label{fig_dispersion_Vs_snr} The power-constrained Rayleigh fading dispersion vs. the unconstrained dispersion.}
\end{figure}

\begin{figure}[htp]
\center{\includegraphics[width=1\columnwidth]{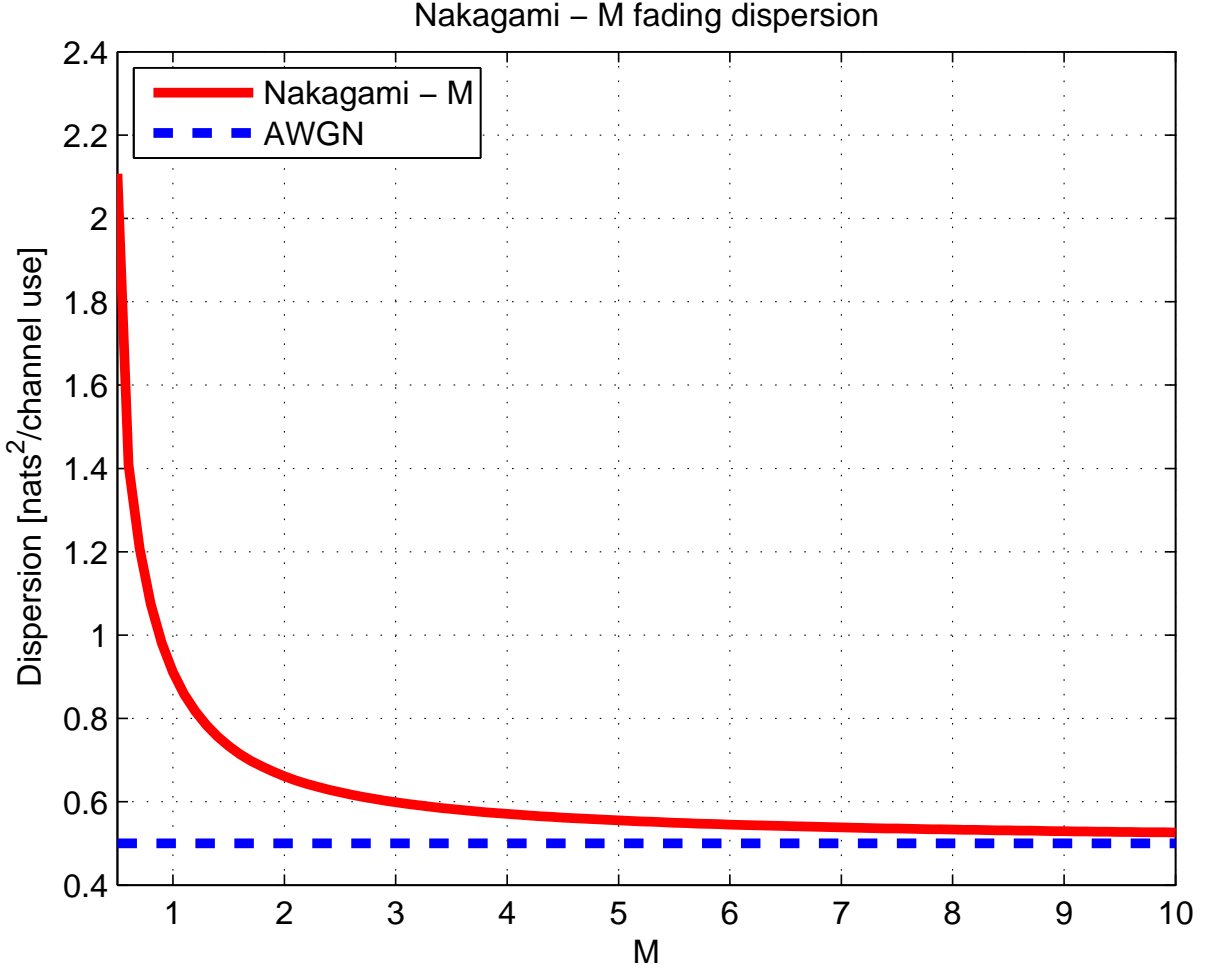}}
\caption{\label{fig_Nakagami_m_dispersion} The IC dispersion of Nakagami - M fading channel converges to the dispersion of the AWGN channel.}
\end{figure}

In Fig. \ref{fig_dispersion_Vs_snr} we can see the power constrained channel dispersion rate of convergence to the unconstrained channel dispersion limit, with growing SNRs, at the popular Rayleigh fading channel. In Fig. \ref{fig_Nakagami_m_dispersion} we can see the unconstrained channel dispersion for the Nakagami-$m$ fading, for various values of $m$. It can be seen that when $m\rightarrow\infty$ the dispersion converges to the unconstrained AWGN channel dispersion $\frac{1}{2}$, as expected.
\section{Main Result}
\label{sec_main_result}
\begin{thm}
\label{thm_main_result}
Let $\epsilon$ be a given, fixed, error probability. Denote by $\delta^*(n,\epsilon)$ the optimal NLD for which there exists an $n$-dimensional infinite constellation with average error probability at most $\epsilon$. Then, for any regular fading distribution of $H$, as $n$ grows,
\begin{equation}
\label{eq_main_result}
\delta^*(n,\epsilon) = \delta^* - \sqrt{\frac{V}{n}}Q^{-1}(\epsilon) + O\left(\frac{\ln(n)}{n}\right),
\end{equation}
where,
\begin{align}
\delta^* &\triangleq E\left\{ \delta(H) \right\} = E\left\{ \frac{1}{2}\ln\left(\frac{H^2}{2\pi e\sigma^2}\right) \right\}
\\ V     &\triangleq \frac{1}{2} + Var(\delta(H)) = \frac{1}{2} + Var\left(\frac{1}{2}\ln(H^2)\right)
\end{align}
noting that
\begin{equation}
\delta(H) \triangleq \frac{1}{2}\ln\left(\frac{H^2}{2\pi e\sigma^2}\right).
\end{equation}
\end{thm}
\begin{cor}
The highest achievable NLD with arbitrary small error probability, over the unconstraint fast fading channel with available CSI at the receiver, is given by
\begin{equation}
\delta^* \triangleq E\left\{ \frac{1}{2}\ln\left(\frac{H^2}{2\pi e\sigma^2}\right) \right\}.
\end{equation}
\end{cor}
\begin{IEEEproof}
By taking the limit $n\rightarrow\infty$ in \eqref{eq_main_result} we get the desired result (for any $0 < \epsilon < 1$).
\end{IEEEproof}

Moreover, by Jensen's inequality and the concavity of the logarithm function, we can derive the following result:
\begin{equation}
\delta^* \triangleq E\left\{ \frac{1}{2}\ln\left(\frac{H^2}{2\pi e\sigma^2}\right) \right\}
\leq \frac{1}{2}\ln\left( E\left\{\frac{H^2}{2\pi e\sigma^2}\right\} \right)
= \frac{1}{2}\ln\left(\frac{1}{2\pi e\sigma^2}\right).
\end{equation}
This proves that in the AWGN channel the Poltyrev's capacity is greater than its equivalent in the fast fading channel (with the same noise variance $\sigma^2$). This loss, relative to the AWGN channel, is given exactly by $-E\left\{\ln(H)\right\}$ in nats. Alternatively, this loss can be measured as the ratio between the highest noise variance that is tolerable in each channel model. It is easy to show that this ratio is given by $e^{-2E\left\{\ln(H)\right\}}$ in linear scale, or by $-8.6859E\left\{\ln(H)\right\}$ in dB. For example, this loss equals approximately 0.288 nats or 2.5 dB in the Rayleigh fading channel.
\subsection{The Sphere Packing Bound}
In this section we will prove the following \emph{sphere packing bound} for any IC $S$ with NLD $\delta$:
\begin{equation}
\label{eq_general_SPB}
P_e\left(S\right) \geq P_e^{\text{SB}}\left(\delta\right) \triangleq Pr\left\{\left\|\mathbf{z}\right\|^2 \geq e^{-2\delta}\left(\frac{\det(\mathbf{H})}{V_n}\right)^{\frac{2}{n}} \right\}.
\end{equation}

First, we will focus on IC where all the Voronoi cells have equal volume $V$.
In the receiver, given the CSI $\mathbf{H}$, we get an IC with Voronoi cell volume that equals $V_{\text{rc}} = V\cdot\det(\mathbf{H}) = V\Pi_{i=1}^{n}{H_i}$.
By the \emph{equivalent sphere} argument \cite{Poltyrev}\cite{Tarokh}, the probability that the noise leaves the Voronoi cell in the receiver is lower bounded by the probability to leave a sphere of the same volume:
\begin{equation}
\label{eq_SPB}
P_e\left(S\right) \geq Pr\left\{\left\|\mathbf{z}\right\|^2 \geq r_{\text{eff}}^{2}(\mathbf{H}) \right\},
\end{equation}
where
\begin{equation}
\label{eq_equivalent_sphere}
V_nr_{\text{eff}}^{n}(\mathbf{H}) \triangleq V_{\text{rc}}
\end{equation}
and
\begin{equation}
\label{eq_sphere_vol_coef}
V_n = \frac{\pi^{n/2}}{\frac{n}{2}\Gamma\left(\frac{n}{2}\right)}.
\end{equation}

Combining \eqref{eq_SPB}, \eqref{eq_equivalent_sphere}, \eqref{eq_sphere_vol_coef} with the definition of $\delta = -\frac{1}{n}\ln(V)$ we get:
\begin{equation}
P_e\left(S\right) \geq Pr\left\{\left\|\mathbf{z}\right\|^2 \geq e^{-2\delta}\left(\frac{\det(\mathbf{H})}{V_n}\right)^{\frac{2}{n}} \right\} \triangleq P_e^{\text{SB}}\left(\delta\right).
\end{equation}

Now we will extend the correctness of the bound to any IC with bounded Voronoi cells volume (\emph{regular IC's}).
\begin{defn}
(Regular IC's): An IC S is called regular if there exists a radius $r_0 > 0$, s.t. for all $s \in S$, the Voronoi cell $W(s)$ is contained in $\emph{\text{Ball}}(s,r_0) \triangleq \left\{ \mathbf{x} \in \mathbb{R}^n ~ s.t. ~\|\mathbf{x} - s\| < r_0 \right\}$.
\end{defn}

For $s \in S$, denote by $v(s)$ the volume of the Voronoi cell of $s$, and denote by $V\left(S\right)$ the average Voronoi cell volume of $S$. Then, by definition
\begin{equation}
V\left(S\right) \triangleq \liminf_{a \rightarrow \infty} E_{S,a}\left\{v(s)\right\} = \liminf_{a \rightarrow \infty} \frac{1}{M(S,a)} \sum_{s \in S\bigcap\text{Cb}(a)}{v(s)}.
\end{equation}
It is easy to verify that for any regular IC, the density is given by $\gamma = \frac{1}{V\left(S\right)}$.

Clearly, for any given $\mathbf{H}$, the receiver IC is also regular. Hence, in the same manner, we can define the receiver's average Voronoi cell volume of $S_{\mathbf{H}}$ by
\begin{equation}
V\left(S_{\mathbf{H}}\right) \triangleq \liminf_{a \rightarrow \infty} E_{S,a|\mathbf{H}}\left\{v(s_{\text{rc}})\right\}.
\end{equation}
The density at the receiver is given by $\gamma_{\text{rc}} = \frac{1}{V\left(S_{\mathbf{H}}\right)} = \frac{\gamma}{\det\left(\mathbf{H}\right)}$.

To prove the sphere bound for regular IC's it is desirable for the clarity of the proof to denote by $\text{SPB}\left(v|\mathbf{H}\right)$, the probability that the noise vector $\mathbf{z}$ leaves a sphere of volume $v$ given the CSI $\mathbf{H}$. With this notation,
\begin{equation}
P_e\left(s_{\text{rc}}|\mathbf{H}\right) \geq \text{SPB}\left(v\left(s_{\text{rc}}\right) | \mathbf{H} \right)
= Pr\left\{\left\|\mathbf{z}\right\|^2 \geq \left(\frac{v\left(s_{\text{rc}}\right)}{V_n}\right)^{\frac{2}{n}} \Big| \mathbf{H} \right\}
\end{equation}
for any $s_{\text{rc}} \in S_{\mathbf{H}}$.

\begin{thm}
For any regular IC $S$ with NLD $\delta$, the average error probability is lower bounded by the following sphere packing bound
\begin{equation}
P_e\left(S\right) \geq P_e^{\text{SB}}\left(\delta\right).
\end{equation}
\end{thm}
\begin{IEEEproof}
By definition the average error probability is given by
\begin{align}
P_e\left(S\right) &\triangleq E\left\{ \limsup_{a \rightarrow \infty} E_{S,a|\mathbf{H}}\left\{ P_e\left(s_{\text{rc}}|\mathbf{H}\right) \right\} \right\}
\\                \label{align_SPB_regular_IC_step_1}
                  &\geq       E\left\{ \limsup_{a \rightarrow \infty} E_{S,a|\mathbf{H}}\left\{ \text{SPB}\left(v\left(s_{\text{rc}}\right) | \mathbf{H} \right) \right\} \right\}
\\                \label{align_SPB_regular_IC_step_2}
                  &\geq       E\left\{ \limsup_{a \rightarrow \infty}  \text{SPB}\left(E_{S,a|\mathbf{H}}\left\{v\left(s_{\text{rc}}\right)\right\} | \mathbf{H} \right)  \right\}
\\                \label{align_SPB_regular_IC_step_3}
                  &=          E\left\{ \text{SPB}\left(\limsup_{a \rightarrow \infty} E_{S,a|\mathbf{H}}\left\{v\left(s_{\text{rc}}\right)\right\} | \mathbf{H} \right)  \right\}
\\                &=          E\left\{ \text{SPB}\left(V\left(S_{\mathbf{H}}\right) | \mathbf{H} \right)  \right\}
\\                &=          Pr\left\{\left\|\mathbf{z}\right\|^2 \geq \left(\frac{V\left(S_{\mathbf{H}}\right)}{V_n}\right)^{\frac{2}{n}} \right\}
\\                &=          Pr\left\{\left\|\mathbf{z}\right\|^2 \geq e^{-2\delta}\left(\frac{\det(\mathbf{H})}{V_n}\right)^{\frac{2}{n}} \right\}
                  \triangleq P_e^{\text{SB}}\left(\delta\right)
\end{align}
where \eqref{align_SPB_regular_IC_step_1} follows from the sphere packing bound for each $s_{\text{rc}} \in S_{\mathbf{H}}$, \eqref{align_SPB_regular_IC_step_2} follows from Jensen's inequality and the convexity of the function $\text{SPB}\left(v|\mathbf{H}\right)$ in $v$ and \eqref{align_SPB_regular_IC_step_3} follows from the fact that $\text{SPB}\left(v|\mathbf{H}\right)$ is monotone decreasing and a continuous function of $v$. All the next steps are trivial.
\end{IEEEproof}

In the next theorem we will extend the correctness of the sphere packing bound for any IC. This includes IC's with unbounded Voronoi's cells and IC's with density which oscillates with the cube size $a$ (i.e. only the limsup exists in the definition of $\gamma$). The proof is based on a very similar regularization process as done in \cite[Lemma 1]{Ingber} for AWGN channels. Here, in the fading channel case, we will need to separate from the analysis all the ``strong'' fading channel realizations, which are formally defined in the following, and use the regularization process only for the rest of the ``weak'' fading realizations. By showing that the ``strong'' fading realizations in regular fading distributions are an arbitrarily small fraction of the whole realizations space, we will complete the proof of the bound.
\begin{defn}
($\xi$ - strong fading realization): Let us denote by $\mathrm{H} = \emph{\text{diag}}\{h_1, \dots, h_n\}$ a fading channel realization drawn from a regular fading distribution of the random fading matrix $\mathbf{H}$. For a given $\xi > 0$, let us define a fading threshold $h^*_{\min}(\xi)$ as the solution of $Pr\{H_{\min} \leq h^*_{\min} \} = \xi$, where $H_{\min} \triangleq \min(H_1, \dots, H_n)$. If $h_{\min} \triangleq \min(h_1, \dots, h_n) \leq h^*_{\min}(\xi)$ then $\mathrm{H}$ is called a $\xi$ - strong fading channel realization.
\end{defn}
\begin{lem}
\label{lem_regularixation}
(Regularization): Given the fading channel realization $\mathbf{H}$, let $S_{\mathbf{H}}$ be an IC with density $\gamma_{\emph{\text{rc}}}\left({\mathbf{H}}\right)$ and average error probability $P_e\left(S_{\mathbf{H}}\right) = \epsilon(\mathbf{H})$. For any $\xi > 0$, if $\mathbf{H}$ is not a $\xi$ - strong fading realization then there exists a regular IC, denoted by $S^{'}_{\mathbf{H}}$, with density $\gamma^{'}_{\emph{\text{rc}}}\left({\mathbf{H}}\right) \geq \gamma_{\emph{\text{rc}}}\left({\mathbf{H}}\right)/(1 + \xi)$ and average error probability $P_e\left(S^{'}_{\mathbf{H}}\right) \leq \epsilon(\mathbf{H})(1 + \xi)$.
\end{lem}
\begin{IEEEproof}
See Appendix \ref{app_proof_regularization_lemma}.
\end{IEEEproof}
\begin{thm}
\label{thm_sphere_packing_lower_bound}
For any IC $S$ with NLD $\delta$, the average error probability is lower bounded by the following sphere packing bound:
\begin{equation}
P_e\left(S\right) \geq P_e^{\text{SB}}\left(\delta\right).
\end{equation}
\end{thm}
\begin{IEEEproof}
For a given $\mathbf{H}$, denote the receiver IC by $S_{\mathbf{H}}$. For any $\xi > 0$, by the \emph{regularization lemma}, if $\mathbf{H}$ is not a \emph{$\xi$ - strong fading realization}, then there exists a regular IC, denoted by $S^{'}_{\mathbf{H}}$, with density
\begin{equation}
\gamma^{'}_{\text{rc}}\left(\mathbf{H}\right) \geq \gamma_{\text{rc}}\left(\mathbf{H}\right)/\left(1+\xi\right)= \frac{\gamma}{\left(1+\xi\right)}\cdot\frac{1}{\det\left({\mathbf{H}}\right)}
\end{equation}
and average error probability
\begin{equation}
P_e\left(S^{'}_{\mathbf{H}}\right) \leq P_e\left(S_{\mathbf{H}}\right)\left(1+\xi\right),
\end{equation}
where $\gamma = e^{n\delta}$. Moreover, by the $\xi$ - strong fading definition $Pr\left\{ H_{\text{min}} \leq h^*_{\text{min}} \right\} = \xi$. Following this, we can derive the inequalities below:
\begin{align}
\left(1+\xi\right)P_e\left(S\right) &= E\left\{ \left(1+\xi\right)P_e\left(S_{\mathbf{H}}\right) \right\}
\\ \label{align_spb_step_1}
   &\geq E\left\{ \left(1+\xi\right)P_e\left(S_{\mathbf{H}}\right) \cdot 1_{\left\{ H_{\text{min}} > h^*_{\text{min}} \right\}} \right\}
\\ \label{align_spb_step_2}
   &\geq E\left\{ P_e\left(S^{'}_{\mathbf{H}}\right) \cdot 1_{\left\{ H_{\text{min}} > h^*_{\text{min}} \right\}} \right\}
\\ \label{align_spb_step_3}
   &\geq E\left\{ \text{SPB}\left( {\gamma^{'-1}_{\text{rc}}} \Big| \mathbf{H} \right) \cdot 1_{\left\{ H_{\text{min}} > h^*_{\text{min}} \right\}} \right\}
\\ \label{align_spb_step_4}
   &\geq E\left\{ \text{SPB}\left( {\gamma}^{-1}\det\left(\mathbf{H}\right)\left(1+\xi\right) \Big| \mathbf{H} \right) \cdot 1_{\left\{ H_{\text{min}} > h^*_{\text{min}} \right\}} \right\}
\\ \label{align_spb_step_5}
   &=    E\left\{ \text{SPB}\left( {\gamma}^{-1}\det\left(\mathbf{H}\right)\left(1+\xi\right) \Big| \mathbf{H} \right) \cdot \left(1 - 1_{\left\{ H_{\text{min}} \leq h^*_{\text{min}} \right\}} \right) \right\}
\\ \label{align_spb_step_6}
   &\geq E\left\{ \text{SPB}\left( {\gamma}^{-1}\det\left(\mathbf{H}\right)\left(1+\xi\right) \Big| \mathbf{H} \right) \right\} - Pr\left\{ H_{\text{min}} \leq h^*_{\text{min}} \right\}
\\ \label{align_spb_step_7}
   &=    E\left\{ \text{SPB}\left( {\gamma}^{-1}\det\left(\mathbf{H}\right)\left(1+\xi\right) \Big| \mathbf{H} \right) \right\} - \xi,
\end{align}
where \eqref{align_spb_step_3} follows from the regularity of $S^{'}_{\mathbf{H}}$, \eqref{align_spb_step_4} is due to the fact that $\text{SPB}\left(\cdot|\mathbf{H}\right)$ is a monotone decreasing function and \eqref{align_spb_step_6} is due to $\text{SPB}\left(\cdot|\mathbf{H}\right) \leq 1$.

Equivalently, we get the following:
\begin{equation}
P_e\left(S\right) \geq E\left\{ \frac{ \text{SPB}\left( {\gamma}^{-1}\det\left(\mathbf{H}\right)\left(1+\xi\right) \Big| \mathbf{H} \right)}{ 1 + \xi } - \frac{\xi}{1 + \xi} \right\}
\end{equation}
for all $\xi > 0$. Since $\text{SPB}(\cdot | \mathbf{H})$ is a continuous function we can take the limit $\xi \rightarrow 0$ (meaning implicitly that the ``strong'' fading realizations are an arbitrarily small fraction of the whole realizations space in regular fading distribution) and get the sphere packing lower bound:
\begin{align}
P_e\left(S\right) &\geq E\left\{ \text{SPB}\left( e^{-n\delta}\det\left(\mathbf{H}\right)\Big| \mathbf{H} \right) \right\}
\\                &=          Pr\left\{\left\|\mathbf{z}\right\|^2 \geq e^{-2\delta}\left(\frac{\det(\mathbf{H})}{V_n}\right)^{\frac{2}{n}} \right\} \triangleq P_e^{\text{SB}}\left(\delta\right).
\end{align}
\end{IEEEproof}

By taking the fading matrix $\mathbf{H}$ to be equal constantly to the identity matrix $I_n$, the bound \eqref{eq_general_SPB} coincides with the sphere packing bound of the unconstrained AWGN channel, which is given by $Pr\left\{\left\|\mathbf{z}\right\| \geq V_n^{-\frac{1}{n}}e^{-\delta} \right\}$. Although this one dimensional integral is hard to evaluate analytically for general $n$, Ingber et al. derived in \cite{Ingber} an easy to evaluate and very tight analytical bounds for it. These bounds coincide with the sphere packing bound's error exponent, derived by Poltyrev in \cite{Poltyrev}, for asymptotic $n$. Moreover, Tarokh et al. represented this integral in \cite{Tarokh} as a sum of $n/2$ elements, which helps in numerical evaluation of the bound. In contrast, in the case of fading channel the sphere packing bound \eqref{eq_general_SPB} is an $n+1$ dimensional integral, which is extremely hard to evaluate both numerically and analytically. Nevertheless, in the asymptotic case, this bound can be approximated by normal distribution according to the central limit theorem, which helps us to prove the converse part of our main result.
\subsection{Proof of Converse Part}
Assume a transmission of IC $S$ with NLD $\delta$ over the fading channel. By the \emph{sphere packing lower bound} of Theorem \ref{thm_sphere_packing_lower_bound},
\begin{equation}
\label{eq_SPB_step1}
P_e \geq P_e^{\text{SB}}\left(\delta\right) = Pr\left\{\left\|\mathbf{z}\right\|^2 \geq e^{-2\delta}\left(\frac{\det(\mathbf{H})}{V_n}\right)^{\frac{2}{n}} \right\}.
\end{equation}

In \cite{Ingber} Ingber et al. proved the converse part of the dispersion analysis, in the unconstrained AWGN channel, by approximating the distribution of $\left\|\mathbf{z}\right\|^2 = \sum_{i=1}^{n}{Z_i^2}$ by a normal distribution using the Berry-Esseen lemma (see Lemma \ref{lem_Berry_Esseen}) for sum of i.i.d RVs. Here, we cannot use the same analysis due to the fact that $\mathbf{H}$ is also random. By taking the logarithm and rearranging of the inequality in the argument of \eqref{eq_SPB_step1} we get:
\begin{align}
\label{align_SPB_step2}
P_e \geq Pr &\Bigg\{ \frac{ \ln\left(\left\|\mathbf{z}\right\|^2\right) - \ln(n\sigma^2) }{\sqrt{\frac{2}{n}}} - \sqrt{\frac{2}{n}}\sum_{i=1}^{n}{\left(\ln(H_i) - E\{\ln(H)\}\right)}
\\ &\geq \sqrt{2n}\left( E\left\{\frac{1}{2}\ln\left(\frac{H^2}{n\sigma^2}\right)\right\} - \delta - \frac{\ln(V_n)}{n} \right) \Bigg\}.
\end{align}

For simplicity, let us define $Y_n \triangleq  \frac{ \ln\left(\left\|\mathbf{z}\right\|^2\right) - \ln\left(n\sigma^2\right) }{\sqrt{\frac{2}{n}}}$,
$S_n \triangleq \frac{\sum_{i=1}^{n}X_i}{\sqrt{n}}$ where
$X_i \triangleq \frac{\ln(H_i)-E\{\ln(H)\}}{\sqrt{Var(\delta(H))}}$ (for $i=1,..,n$) and $\zeta_n \triangleq \frac{1}{\sqrt{2}}Y_n - \sqrt{Var(\delta(H))}S_n$ to get:
\begin{equation}
\label{eq_SPB_step3}
P_e \geq
Pr\left\{ \zeta_n \geq \zeta \right\},
\end{equation}
where $\zeta \triangleq \sqrt{n}\left( E\left\{\frac{1}{2}\ln\left(\frac{H^2}{n\sigma^2}\right)\right\} - \delta - \frac{\ln(V_n)}{n} \right)$.

Although $\zeta_n$ is a sum of $n+1$ independent RVs, and despite of the existence of expansions for the Berry-Essen Lemma for a sum of independent RVs with varying distributions, in the standard derivation of these expansions it is assumed that all the RVs' variances are of the same order (see \cite[pp. 542-548]{Feller} for details). Here, $Var(Y_n) = O(1)$ (see Lemma \ref{lem_ln_chi2_pdf}) and $Var\left(\frac{X_i}{\sqrt{n}}\right) = O\left(\frac{1}{n}\right)$. Hence, a more careful analysis should be done for proving that the distribution of $\zeta_n$ is approximately normal. The following three lemmas allow it. By Lemma \ref{lem_ln_chi2_pdf} and by Lemma \ref{lem_Berry_Esseen} we prove that the PDF of $Y_n$ and the CDF of $S_n$ are approximately normal for large enough $n$, respectively. Finally by Lemma \ref{lem_sum_of_almost_normal_RVs} we prove that the distribution of a sum of two independent RVs, each of which has an approximately normal distribution, is also approximately normal. Therefore, the distribution of $\zeta_n$ is also approximately normal for large enough $n$.
\begin{lem}
\label{lem_ln_chi2_pdf}
(Log of chi square distribution) Let $Y_n \triangleq \frac{\ln\left(X\right)-\ln(n)}{\sqrt{\frac{2}{n}}}$, where $X \sim \chi^2_n$. Then
\begin{equation}
\label{eq_Yn_PDF}
f_{Y_n}(y) = \frac{ (\frac{n}{2})^{\frac{n-1}{2}}}{\Gamma(\frac{n}{2})}e^{\sqrt{\frac{n}{2}}y - \frac{n}{2}e^{\sqrt{\frac{2}{n}}y}},
\end{equation}
and for large enough $n$:
\begin{equation}
\label{eq_Yn_asymptotic_PDF}
f_{Y_n}(y) = N(0,1) + e_n(y)
\\~s.t. \int_{-\infty}^{\infty}|e_n(y)|dy = O\left(\frac{1}{\sqrt{n}}\right),
\end{equation}
where $N(0,1)$ is the standard normal distribution's PDF.
\end{lem}
\begin{IEEEproof}
See Appendix \ref{app_proof_lemma_ln_chi2_density_moments}.
\end{IEEEproof}
\begin{lem}
\label{lem_Berry_Esseen}
(Berry-Esseen) Let $X_1,X_2,\dots,X_n$ be $n$ i.i.d. random variables with mean, variance and third absolute moments that equal $\mu = E\{X_i\}, \sigma^2 = Var(X_i)$ and $\rho_3=E\{|X_i-\mu|^3\}$, respectively, for $i=1,\dots,n$. If the third absolute moment exists, then for all $-\infty < s < \infty$ and $n$,
\begin{equation}
\Big| F_{S_n}(s) - F_{N(0,1)}(s) \Big| \leq \frac{6\rho_3}{\sqrt{n}\sigma^3},
\end{equation}
where $S_n\triangleq\frac{\sum_{i=1}^{n}(X_i - \mu)}{\sqrt{n}\sigma}$ and $F_{N(0,1)}(\cdot)$ is the standard normal distribution's CDF.
\end{lem}
\begin{IEEEproof}
See Theorem 1 (Berry-Esseen for sum of i.i.d. RVs) in \cite[pp. 542]{Feller}.
\end{IEEEproof}
\begin{lem}
\label{lem_sum_of_almost_normal_RVs}
(Sum of two almost normal RVs) Suppose that $X_1$ and $X_2$ are two independent random variables s.t. the PDF of $X_1$ is given by
\begin{equation}
f_{X_1}(x_1) = N(0, \sigma_1^2) + e_n(y)
\\~s.t. \int_{-\infty}^{\infty}|e_n(y)|dy = O\left(\frac{1}{\sqrt{n}}\right), \nonumber
\end{equation}
and the CDF of $X_2$ is given by
\begin{equation}
F_{X_2}(x_2) = F_{N(0,\sigma_2^2)}(x_2) + O\left(\frac{1}{\sqrt{n}}\right). \nonumber
\end{equation}

Let $Y \triangleq X_1 + X_2$, then the following holds:
\begin{equation}
\label{eq_sum_of_almost_normal_CDF}
F_Y(y) = F_{N(0,\sigma_y^2)}(y) + O\left(\frac{1}{\sqrt{n}}\right),
\end{equation}
where $\sigma_y^2 \triangleq \sigma_1^2 + \sigma_2^2$.
\end{lem}
\begin{IEEEproof}
See Appendix \ref{app_proof_lemma_sum_of_almost_normal_RVs}.
\end{IEEEproof}

Combining Lemmas \ref{lem_ln_chi2_pdf}, \ref{lem_Berry_Esseen} and \ref{lem_sum_of_almost_normal_RVs} we get:
\begin{equation}
\label{eq_SPB_step4}
P_e \geq Q\left(\frac{\zeta}{\sqrt{V}}\right) - O\left(\frac{1}{\sqrt{n}}\right).
\end{equation}
By Stirling approximation for the Gamma function, $V_n$ can be approximated as
\begin{equation}
\frac{\ln(V_n)}{n} = \frac{1}{2}\ln\left(\frac{2\pi e}{n}\right) - \frac{1}{2n}\ln(n) + O\left(\frac{1}{n}\right)
\end{equation}
and hence we get:
\begin{equation}
\label{eq_zeta}
\zeta = \sqrt{n}\left( \delta^* - \delta + \frac{1}{2n}\ln(n) + O\left(\frac{1}{n}\right) \right).
\end{equation}
The assignment of \eqref{eq_zeta} in \eqref{eq_SPB_step4} gives us:
\begin{equation}
\label{eq_SPB_step5}
\epsilon \geq P_e \geq
Q\left(\frac{ \delta^* - \delta + \frac{1}{2n}\ln(n) + O\left(\frac{1}{n}\right) }{ \sqrt{ \frac{V}{n} } }\right) - O\left(\frac{1}{\sqrt{n}}\right).
\end{equation}
Taking $Q^{-1}(\cdot)$ from both sides of \eqref{eq_SPB_step5} gives us:
\begin{equation}
\label{eq_SPB_step6}
\delta \leq \delta^* - \sqrt{ \frac{V}{n} }Q^{-1}\left(\epsilon + O\left(\frac{1}{\sqrt{n}}\right)\right) + \frac{1}{2n}\ln(n) + O\left(\frac{1}{n}\right).
\end{equation}

By Taylor approximation (around $\epsilon$) $  Q^{-1}\left(\epsilon + O\left(\frac{1}{\sqrt{n}}\right)\right) = Q^{-1}(\epsilon) + O\left(\frac{1}{\sqrt{n}}\right)$, which gives us the desired result:
\begin{equation}
\label{eq_SPB_step7}
\delta \leq \delta^* - \sqrt{ \frac{V}{n} }Q^{-1}(\epsilon) + \frac{1}{2n}\ln(n) + O\left(\frac{1}{n}\right).
\end{equation}
\subsection{Dependence Testing Bound}
In this section we will extend Polyanskiy's \emph{Dependence Testing Bound} (see Theorems 17 and 18 in \cite{Polyanskiy}) for the case of fast fading channel with available CSI at the receiver. In \cite{Polyanskiy} the DT bound was used to prove the dispersion analysis for DMC, or more precisely, for memoryless channels without a power constraint (or any other constraint on the channel input). Here, the channel input doesn't have any restriction, and hence we can use the DT bound to prove the direct part of our main result.
\begin{thm}
\label{thm_DT_bound}
(DT bound) For any input distribution $f_X(x)$ on $\mathbb{R}$, there exists a code with $M$ codewords and an average error probability over the fast fading channel, with available CSI at the receiver, not exceeding
\begin{equation}
\label{eq_DT_bound}
P_e \leq Pr\left\{ i(\mathbf{x};\mathbf{y},\mathbf{H}) \leq \ln\left(\frac{M-1}{2}\right) \right\}
    + \frac{M-1}{2} Pr\left\{i(\mathbf{x};\bar{\mathbf{y}},\mathbf{H}) > \ln\left(\frac{M-1}{2}\right)\right\},
\end{equation}
or equivalently,
\begin{align}
\begin{aligned}
\label{align_DT_bound_equivalence}
P_e &\leq E\left\{e^{-\left[i(\mathbf{x};\mathbf{y},\mathbf{H}) - \ln\left(\frac{M-1}{2}\right)\right]^+}\right\}
\\ &= Pr\left\{ i\left(\mathbf{x};\mathbf{y},\mathbf{H}\right) \leq \ln\left(\frac{M-1}{2}\right) \right\}
+ \frac{M-1}{2} E\left\{ e^{-i\left(\mathbf{x};\mathbf{y},\mathbf{H}\right)}1_{\left\{i\left(\mathbf{x};\mathbf{y},\mathbf{H}\right)>\ln\left(\frac{M-1}{2}\right)\right\}} \right\},
\end{aligned}
\end{align}
where $f_{XY\bar{Y}H}(x,y,\bar{y},h) = f_X(x)f_{Y|X,H}(y|x,h)f_{Y|H}(\bar{y}|h)f_H(h)$ is the marginal joint distribution of all the random vectors arising above and $i(\mathbf{x};\mathbf{y},\mathbf{H}) \triangleq \ln\left( \frac{f(\mathbf{x}, \mathbf{y}, \mathbf{H})}{f(\mathbf{x})f(\mathbf{y}, \mathbf{H})} \right)$.
\end{thm}
\begin{IEEEproof}
The proof is based on Shannon's random coding technique and on a suboptimal decoder. For a given input distribution $f_X(x)$ , let us define the following deterministic function:
\begin{equation}
g_{\mathbf{x}}\left(\mathbf{y},\mathbf{H}\right) = 1_{\left\{i(\mathbf{x};\mathbf{y},\mathbf{H}) > \ln\left(\frac{M-1}{2}\right)\right\}}.
\end{equation}

For a given codebook $C = \left\{c_1, \dots, c_M\right\}$, the decoder computes the $M$ values of $g_{c_j}\left(\mathbf{y},\mathbf{H}\right)$ for the given channel output $\left(\mathbf{y},\mathbf{H}\right)$ and returns the lowest index $j$ for which $g_{c_j}\left(\mathbf{y},\mathbf{H}\right) = 1$, or declares an error if there is no such index. Hence, the error probability, given that $\mathbf{x} = c_j$ was transmitted, is given by:
\begin{align}
\label{align_DT_bound_step_1}
\begin{aligned}
Pr\left\{ \{g_{c_j}\left(\mathbf{y},\mathbf{H}\right) = 0\} \bigcup_{i<j} \{g_{c_i}\left(\mathbf{y},\mathbf{H}\right) = 1\} | ~\mathbf{x} = c_j \right\} &\leq
  Pr\left\{ i(c_j;\mathbf{y},\mathbf{H}) \leq \ln\left(\frac{M-1}{2}\right) | ~\mathbf{x} = c_j \right\}
\\ &+ \sum_{i < j}Pr\left\{ i(c_i;\bar{\mathbf{y}},\mathbf{H}) > \ln\left(\frac{M-1}{2}\right) | ~\mathbf{x} = c_j \right\},
\end{aligned}
\end{align}
where the RHS of \eqref{align_DT_bound_step_1} is obtained by using the union bound and the definition of $\bar{\mathbf{y}}$ as a random vector which is independent of $\mathbf{x}$ given $\mathbf{H}$ and has the same distribution as $\mathbf{y}$ given $\mathbf{H}$.

Let us define the ensemble of the codebooks of size M, that every codeword's component in it is drawn independently of each other by $f_X(x)$. Averaging \eqref{align_DT_bound_step_1} over this ensemble and over the $M$ equiprobable codewords we obtain
\begin{align}
\begin{aligned}
P_e &\leq Pr\left\{ i(\mathbf{x};\mathbf{y},\mathbf{H}) \leq \ln\left(\frac{M-1}{2}\right) \right\}
    + \sum_{j=1}^{M}{ \frac{j-1}{M}Pr\left\{i(\mathbf{x};\bar{\mathbf{y}},\mathbf{H}) > \ln\left(\frac{M-1}{2}\right)\right\} },
\end{aligned}
\end{align}
which completes the proof of the existence of a code with $M$ codewords whose average error probability is upper bounded by \eqref{eq_DT_bound}.

Now we turn to prove the equivalent bound \eqref{align_DT_bound_equivalence} of the theorem. For any positive $\gamma$ the following identities hold:
\begin{align}
E\left\{e^{-\left[i(\mathbf{x};\mathbf{y},\mathbf{H}) - \ln(\gamma)\right]^+}\right\}
   &= E\left\{ 1_{\left\{i\left(\mathbf{x};\mathbf{y},\mathbf{H}\right) \leq \ln(\gamma) \right\}}
   + \gamma e^{-i\left(\mathbf{x};\mathbf{y},\mathbf{H}\right)}1_{\left\{i\left(\mathbf{x};\mathbf{y},\mathbf{H}\right)>\ln(\gamma) \right\}} \right\}
\\ &= Pr\left\{ i\left(\mathbf{x};\mathbf{y},\mathbf{H}\right) \leq \ln(\gamma) \right\}
+ \gamma E\left\{ e^{-i\left(\mathbf{x};\mathbf{y},\mathbf{H}\right)}1_{\left\{i\left(\mathbf{x};\mathbf{y},\mathbf{H}\right)>\ln(\gamma) \right\}} \right\}
\\ &= Pr\left\{ i\left(\mathbf{x};\mathbf{y},\mathbf{H}\right) \leq \ln(\gamma) \right\}
+ \gamma E\left\{ \frac{f(\mathbf{x})f(\mathbf{y},\mathbf{H})}{f(\mathbf{x},\mathbf{y},\mathbf{H})} 1_{\left\{i\left(\mathbf{x};\mathbf{y},\mathbf{H}\right)>\ln(\gamma) \right\}} \right\}
\\ &= Pr\left\{ i\left(\mathbf{x};\mathbf{y},\mathbf{H}\right) \leq \ln(\gamma) \right\}
+ \gamma Pr\left\{ i(\mathbf{x};\bar{\mathbf{y}},\mathbf{H}) > \ln(\gamma) \right\}.
\end{align}
By taking $\gamma = \frac{M-1}{2}$ we complete the proof.

It is important to notice that the dependence testing bound is based on a suboptimal decoder which is actually a threshold crossing decoder. The decoder computes $M$ binary hypothesis tests in parallel and declares as the decoded codeword the first one that crosses the threshold $\ln\left(\frac{M-1}{2}\right)$.
\end{IEEEproof}
\subsection{Proof of Direct Part}
For the proof of the direct part, we will first construct an ensemble of finite constellation with $M$ codewords uniformly distributed in an $n$ dimensional cube $\text{Cb}(a)$. Then, using the \emph{Dependence Testing bound} of Theorem \ref{thm_DT_bound} with $f_X(x) = U(-\frac{a}{2},\frac{a}{2})$, we will find a lower bound on $M$ for a FC in such an ensemble, whose error probability is upper bounded by some $\epsilon > 0$. We will denote this lower bound by $M(n,\epsilon,a/\sigma)$. Theorem \ref{thm_DT_bound} also ensures the existence of such a FC that achieves this lower bound. Finally, we will construct an IC by tiling this FC to the whole space $\mathbb{R}^n$, in a way that will preserve the density of codewords and the error probability, asymptotically in the dimension $n$, as in this FC.

To use the DT bound of Theorem \ref{thm_DT_bound}, we need to prove that for some $\gamma$ the following inequality holds:
\begin{align}
\label{align_DT_bound}
P_e \leq Pr\left\{ i\left(\mathbf{x};\mathbf{y},\mathbf{H}\right) \leq \ln(\gamma) \right\} + \gamma E\left\{ e^{-i\left(\mathbf{x};\mathbf{y},\mathbf{H}\right)}1_{\left\{i\left(\mathbf{x};\mathbf{y},\mathbf{H}\right)>\ln(\gamma)\right\}} \right\} \leq \epsilon.
\end{align}

Denote for arbitrary $\tau$
\begin{equation}
\ln(\gamma) = nI(X;Y,H) - \tau\sqrt{nVar(i(X;Y,H))}.
\end{equation}

The information density is a sum of $n$ i.i.d. RVs:
\begin{equation}
i\left(\mathbf{x};\mathbf{y},\mathbf{H}\right) = \sum_{j=1}^{n}{ i(X_j;Y_j,H_j) },
\end{equation}
where $i(X;Y,H) \triangleq \ln\left( \frac{f(Y|H,X)}{f(Y|H)} \right)$ and its moments are given by the following lemma.
\begin{lem}
\label{lem_information_density_moments}
(Information density's moments) If $X\sim U\left(-\frac{a}{2},\frac{a}{2}\right)$ and if the PDF of $H$ is a regular fading distribution, then for large enough $a/\sigma$ and for some positive constant $0<\alpha\leq1$, the moments of the information density $i(X;Y,H)$ are given by:
\begin{enumerate}
  \item $I(X;Y,H) \triangleq E\{i(X;Y,H)\} = E\left\{ \frac{1}{2}\ln\left(\frac{a^2H^2}{2\pi e\sigma^2}\right) \right\} + O\left((\frac{\sigma}{a})^{\alpha}\right)$
  \item $Var(i(X;Y,H)) = \frac{1}{2} + Var(\delta(H)) + O\left((\frac{\sigma}{a})^{\frac{\alpha}{2}}\right)$
  \item $\rho_3 \triangleq E\left\{|i(X;Y,H) - I(X;Y,H)|^3\right\} < \infty$.
\end{enumerate}
\end{lem}
\begin{IEEEproof}
See Appendix \ref{app_proof_lemma_information_density_moments}.
\end{IEEEproof}

According to the Berry-Essen lemma (see Lemma \ref{lem_Berry_Esseen}) for i.i.d. RVs,
\begin{equation}
\label{eq_using_Berry_Esseen_UB}
|Pr\{ i\left(\mathbf{x};\mathbf{y},\mathbf{H}\right) \leq \ln\gamma \} - Q(\tau)| \leq \frac{B(a/\sigma)}{\sqrt{n}}
\end{equation}
where
\begin{equation}
\label{eq_B}
B(a/\sigma) = \frac{6\rho_3}{Var^{\frac{3}{2}}(i(X;Y,H))}.
\end{equation}
For sufficiently large $n$, let
\begin{equation}
\tau = Q^{-1}\left(\epsilon - \left(\frac{2\ln(2)}{\sqrt{2\pi Var(i(X;Y,H))}} + 5B(a/\sigma)\right)\frac{1}{\sqrt{n}} \right).
\end{equation}
Then, from \eqref{eq_using_Berry_Esseen_UB} we obtain
\begin{equation}
\label{eq_part1_UB_for_DT_bound}
Pr\left\{ i\left(\mathbf{x};\mathbf{y},\mathbf{H}\right) \leq \ln(\gamma) \right\} \leq \epsilon - 2\left(\frac{\ln(2)}{\sqrt{2\pi Var(i(X;Y,H))}} + 2B(a/\sigma)\right)\frac{1}{\sqrt{n}}.
\end{equation}
Using Lemma \ref{lem_47_in_Polyanskiy} (see in Appendix \ref{app_lemma_47_in_Polyanskiy}), we get
\begin{equation}
\label{eq_part2_UB_for_DT_bound}
\gamma E\left\{ e^{-i\left(\mathbf{x};\mathbf{y},\mathbf{H}\right)}1_{\left\{i\left(\mathbf{x};\mathbf{y},\mathbf{H}\right)>\ln(\gamma)\right\}} \right\} \leq 2\left(\frac{\ln(2)}{\sqrt{2\pi Var(i(X;Y,H))}} + 2B(a/\sigma)\right)\frac{1}{\sqrt{n}}.
\end{equation}
Summing \eqref{eq_part1_UB_for_DT_bound} and \eqref{eq_part2_UB_for_DT_bound} we prove the inequality \eqref{align_DT_bound}. Hence, by Theorem \ref{thm_DT_bound}, there exists a FC with $M(n,\epsilon,a/\sigma)$ codewords, denoted by $S(n, \epsilon, a/\sigma)$, such that
\begin{align}
\label{align_Direct_step_1}
\begin{aligned}
\ln \left(M(n,\epsilon,a/\sigma)\right) &= \ln(\gamma) + O(1)
\\ &= nI(X;Y,H) - \tau\sqrt{nVar(i(X;Y,H))}  + O(1)
\\ &= nI(X;Y,H) - \sqrt{nVar(i(X;Y,H))}Q^{-1}(\epsilon) + O(1),
\end{aligned}
\end{align}
where the last equality is derived by Taylor approximation for $Q^{-1}\left(\epsilon + O\left(\frac{1}{\sqrt{n}}\right)\right)$ around $\epsilon$.
Let us define the NLD of the FC in $\text{Cb}(a)$ by
\begin{equation}
\label{eq_NLD_in_FC}
\delta(n, \epsilon, a/\sigma) \triangleq \frac{1}{n}\ln\left(\frac{M(n,\epsilon,a/\sigma)}{a^n}\right).
\end{equation}
From \eqref{align_Direct_step_1} we obtain
\begin{align}
\label{align_Direct_step_2}
\delta(n, \epsilon, a/\sigma) = I(X;Y,H) - \ln(a) - \sqrt{\frac{Var(i(X;Y,H))}{n}}Q^{-1}(\epsilon) + O\left(\frac{1}{n}\right).
\end{align}
Note that the results of Lemma \ref{lem_information_density_moments} hold in general for large enough $a$. Specifically, we can choose $a$ to be a monotonic increasing function of $n$ s.t. $\lim_{n \to \infty}a=\infty$, and then the results of Lemma \ref{lem_information_density_moments} will hold for any large enough $n$. Assigning the results of Lemma \ref{lem_information_density_moments} with appropriate choice of $a = a(n)$, we get
\begin{align}
\label{align_Direct_step_3}
\delta(n, \epsilon, a/\sigma) &= \delta^* -
\sqrt{ \frac{V + O\left(\left(\frac{\sigma}{a}\right)^{\frac{\alpha}{2}}\right)}{n} }Q^{-1}(\epsilon)
+ O\left(\frac{1}{n} + \left(\frac{\sigma}{a}\right)^{\alpha}\right).
\end{align}
Using Taylor approximation for large enough $n$,
\begin{equation}
\sqrt{V + O\left(\left(\frac{\sigma}{a}\right)^{\frac{\alpha}{2}}\right)} =
\sqrt{V} + O\left(\left(\frac{\sigma}{a}\right)^{\frac{\alpha}{2}}\right).
\end{equation}
Hence, we get
\begin{align}
\label{align_Direct_step_4}
\delta(n, \epsilon, a/\sigma) = \delta^* - \sqrt{ \frac{V}{n} }Q^{-1}(\epsilon) +
O\left(\frac{1}{n} + \frac{1}{\sqrt{n}}\left(\frac{\sigma}{a}\right)^{\frac{\alpha}{2}} + \left(\frac{\sigma}{a}\right)^{\alpha}\right).
\end{align}

By tiling the FC, denoted by $S(n, \epsilon, a/\sigma)$, to the whole space $\mathbb{R}^n$ and by choosing for example $a(n) = \sigma\cdot n^{2+\frac{2}{\alpha}}$, we can construct an IC (See Appendix \ref{app_tiling} for details) with average error probability which is upper bounded by $\epsilon$ and NLD $\delta(n,\epsilon)$ that satisfies
\begin{equation}
\label{eq_Direct_step_5}
\delta(n, \epsilon) = \delta^* - \sqrt{ \frac{V}{n} }Q^{-1}(\epsilon) + O\left(\frac{1}{n}\right).
\end{equation}
Hence, the optimal NLD $\delta^*(n,\epsilon)$ necessarily satisfies
\begin{align}
\label{align_Direct_step_6}
\delta^*(n, \epsilon) &\geq \delta(n, \epsilon) = \delta^* - \sqrt{ \frac{V}{n} }Q^{-1}(\epsilon) + O\left(\frac{1}{n}\right),
\end{align}
which completes the proof of the direct part.

We can observe that in the case of AWGN, namely $H = 1$ deterministically, our result coincides with the weaker achievability
bound of the dispersion analysis of Ingber et al. in \cite{Ingber}. This weaker bound is based on the suboptimal typicality decoder. The stronger bound in \cite{Ingber}, which is based on the optimal ML decoder, is greater than the typicality bound in $\frac{1}{2n}\ln(n)$. Hence, we conjecture that by using a ML decoder, instead of the suboptimal dependence testing decoder, the achievability bound is, actually, given by:
\begin{equation}
\delta^*(n, \epsilon) \geq \delta^* - \sqrt{ \frac{V}{n} }Q^{-1}(\epsilon) + \frac{1}{2n}\ln(n) + O\left(\frac{1}{n}\right).
\end{equation}
\section{Extension to The Complex Model}
\label{sec_extension_to_the_complex_model}
In this section we will extend our main result to the complex model. First, we will define the complex fading channel model and then we will explain its similarity to the scalar model. Finally, we will give the outline of the proof of the theorem in this setting.

In the complex model, $Y = H\cdot X + Z$ where $X$, $H$ and $Z$ are independent complex RVs. Moreover, $E\left\{|H|^2\right\} = 1$ and $Z \sim CN(0, \sigma^2)$ with i.i.d. real and imaginary
components.

Generally, $H$ is a complex RV, but since in our model the CSI is known at the receiver, we can assume
that $H$ is a real and nonnegative RV, without loss of generality. Hence, the complex model is equivalent to the following two scalar models:
\begin{align}
Y_r = |H|\cdot X_r + Z_r
\\
Y_i = |H|\cdot X_i + Z_i
\end{align}
where, $X = X_r + jX_i$, $Y = Y_r + jY_i$ and $Z = Z_r + jZ_i$.
\begin{thm}
\label{thm_main_result_complex_extension}
Let $\epsilon$ be a given, fixed, error probability. Denote by $\delta_c^*(n,\epsilon)$ the optimal NLD for which there exists an $n$ complex-dimensional infinite constellation with average error probability at most $\epsilon$. Then, for any regular fading distribution of $|H|$, as $n$ grows,
\begin{equation}
\label{eq_main_result_complex}
\delta_c^*(n,\epsilon) = \delta_c^* - \sqrt{\frac{V_c}{n}}Q^{-1}(\epsilon) + O\left(\frac{\ln(n)}{n}\right),
\end{equation}
where,
\begin{align}
\delta_c^* &\triangleq E\left\{\delta_c(H)\right\} = E\left\{ \ln\left(\frac{|H|^2}{\pi e\sigma^2}\right) \right\}
\\ V_c     &\triangleq 1 + Var(\delta_c(H)) = 1 + Var\left(\ln\left(|H|^2\right)\right)
\end{align}
noting that
\begin{equation}
\delta_c(H) \triangleq \ln\left(\frac{|H|^2}{\pi e\sigma^2}\right).
\end{equation}
\end{thm}
\subsection{Proof outline of the direct part}
For the proof of the direct part we need to construct an ensemble of finite constellation with $M$ codewords uniformly distributed in an $n$ complex-dimensional cube $\text{Cb}(a)$. To be more precise, each codeword's component (its real and imaginary parts) in the ensemble is drawn uniformly according to the distribution $U(-\frac{a}{2},\frac{a}{2})$ independently of each other. Then, using the \emph{Dependence Testing bound} of Theorem \ref{thm_DT_bound} over this ensemble and the Berry-Essen Lemma, we can obtain the existence of FC with $M(n,\epsilon,a/\sigma)$ codewords and with an average error probability upper bounded by $\epsilon$, which satisfies the following:
\begin{align}
\label{align_Direct_complex_case_step_1}
\delta_c(n, \epsilon, a/\sigma) \triangleq \ln\left(\frac{M(n,\epsilon,a/\sigma)}{a^{2n}}\right) = I(X;Y,H) - \ln(a^2) - \sqrt{\frac{Var(i(X;Y,H))}{n}}Q^{-1}(\epsilon) + O\left(\frac{1}{n}\right).
\end{align}

In this case the information density is given by
\begin{align}
\begin{aligned}
i(X;Y,H) &= \ln\left( \frac{f(Y|X,H)}{f(Y|H)} \right)
\\       &= \ln\left( \frac{f(Y_r|X_r,|H|)f(Y_i|X_i,|H|)}{f(Y_r||H|)f(Y_i||H|)} \right)
\\       &= i(X_r;Y_r,|H|) + i(X_i;Y_i,|H|).
\end{aligned}
\end{align}
Hence, by equivalent calculations as in Lemma \ref{lem_information_density_moments}, we can obtain
\begin{align}
\begin{aligned}
\label{align_moments_in_complex_case}
I(X;Y,H) &= E\left\{ \ln\left(\frac{a^2|H|^2}{\pi e\sigma^2}\right) \right\} + o(1)
\\
Var(i(X;Y,H)) &= 1 + Var\left(\ln\left(\frac{a^2|H|^2}{\pi e\sigma^2}\right)\right) + o(1),
\end{aligned}
\end{align}
where $o(1)$ converges to zero when $\sigma/a$ tends to zero. Combining \eqref{align_Direct_complex_case_step_1} and \eqref{align_moments_in_complex_case} gives us the following:
\begin{align}
\label{align_Direct_complex_case_step_2}
\delta_c(n, \epsilon, a/\sigma) = \delta_c^* + o(1) - \sqrt{\frac{V_c + o(1)}{n}}Q^{-1}(\epsilon) + O\left(\frac{1}{n}\right).
\end{align}

By tiling this FC to the whole space $\mathbb{C}^n$ we can prove the existence of IC with an average error probability upper bounded by $\epsilon$ and NLD that equals the RHS of \eqref{eq_main_result_complex}. This completes the proof of the direct part.
\subsection{Proof outline of the converse part}
Using the same arguments as in the scalar fading channel model, we can prove that the sphere packing lower bound in the complex model is given by
\begin{equation}
\label{eq_SPB_complex_step1}
P_e \geq P_e^{\text{SB}}\left(\delta_c\right) = Pr\left\{\left\|\mathbf{z}\right\|^2 \geq e^{-\delta_c}\left(\frac{\det(\mathbf{H})^2}{V_{2n}}\right)^{\frac{1}{n}} \right\}
\end{equation}
for any IC $S$ with NLD $\delta_c$, where $\left\|\mathbf{z}\right\|^2 / \sigma^2 \sim \chi^2(2n)$ and $\mathbf{H} = \text{diag}\{{H_1,\dots,H_n}\}$.

Using the same arguments as in the case of the scalar fading model, we can prove that for any $n$ complex-dimensional IC, with NLD $\delta_c$ and average error probability upper bounded by $\epsilon$ over the complex fading channel, the following holds:
\begin{equation}
\label{eq_converse_complex_case}
\delta_c \leq \delta^*_c - \sqrt{ \frac{V_c}{n} }Q^{-1}(\epsilon) + \frac{1}{2n}\ln(n) + O\left(\frac{1}{n}\right),
\end{equation}
which completes the proof of the converse part.
\section{Summary}
\label{sec_summary}
In this paper we derived the dispersion normal approximation for the unconstrained fast fading channel model with perfect CSI at the receiver. We extended the dependence testing bound, derived in \cite{Polyanskiy} for DMCs, and the sphere packing bound, derived in \cite{Ingber} for unconstrained AWGN, to the setting of fast fading channels. By using these extensions (and some normal approximation techniques), we proved the direct and the converse part of our main result, respectively. The connection to the power constrained channel model was also discussed, and it was shown that the unconstrained model can be interpreted as the limit of the power constrained model when the SNR tends to infinity. Finally, we extended the result to the case of a complex fading model.
\begin{appendices}
\section{Proof of the Regularization Lemma}
\label{app_proof_regularization_lemma}
\begin{IEEEproof}[Proof of Lemma \ref{lem_regularixation}]
Fix $\xi > 0$ and consider the receiver's IC $S_{\mathbf{H}}$, where $\mathbf{H}$ is not a \emph{$\xi$ - strong fading realization}. First, we will find large enough $a_*$ s.t. the density of the codewords in $S_{\mathbf{H}} \bigcap \mathbf{H}\cdot\text{Cb}(a_*)$, and the average error probability in transmitting codewords from it, over the AWGN channel, are close enough to $\gamma_{\text{rc}}(\mathbf{H})$ and $\epsilon(\mathbf{H})$. Then we will construct a regular IC by tiling this FC over the whole space $\mathbb{R}^n$. For this IC the desired bounds of the lemma will hold.

By definition we have
\begin{equation}
P_e\left(S_{\mathbf{H}}\right) = P_e(S|\mathbf{H}) = \epsilon(\mathbf{H}) = \limsup_{a \to \infty} \frac{1}{M(S_{\mathbf{H}},a)} \sum_{s_{\text{rc}} \in S_{\mathbf{H}} \bigcap \mathbf{H}\cdot\text{Cb}(a)} P_e(s_{\text{rc}} | \mathbf{H})
\end{equation}
\begin{equation}
\gamma_{\text{rc}}(\mathbf{H}) = \gamma_{\text{rc}} = \limsup_{a \to \infty} \frac{M(S_{\mathbf{H}},a)}{\text{Vol}(\mathbf{H}\cdot\text{Cb}(a))}
                                                    = \limsup_{a \to \infty} \frac{M(S_{\mathbf{H}},a)}{\det(\mathbf{H})a^n}.
\end{equation}

From the existence of the limits above there exists $a_0$ s.t. for every $a > a_0$ the following holds:
\begin{equation}
\label{eq_sup_error_probability}
\sup_{b > a} \frac{1}{M(S_{\mathbf{H}},b)} \sum_{s_{\text{rc}} \in S_{\mathbf{H}} \bigcap \mathbf{H}\cdot\text{Cb}(b)} P_e(s_{\text{rc}} | \mathbf{H}) < \epsilon(\mathbf{H})(1 + \xi/2)
\end{equation}
and
\begin{equation}
\label{eq_sup_density}
\sup_{b > a} \frac{M(S_{\mathbf{H}},b)}{\det(\mathbf{H})b^n} > \frac{\gamma_{\text{rc}}}{\sqrt{1 + \xi}}.
\end{equation}

Define $\Delta$ s.t.
\begin{equation}
2nQ\left( \frac{h^*_{\min}\Delta}{\sigma} \right) = \frac{\xi}{2}\cdot\epsilon(\mathbf{H}),
\end{equation}
and define $a_{\Delta}$ as the solution of
\begin{equation}
\frac{\text{Vol}(\mathbf{H}\cdot\text{Cb}(a_{\Delta} + 2\Delta))}{\text{Vol}(\mathbf{H}\cdot\text{Cb}(a_{\Delta}))}
= \left( \frac{a_{\Delta} + 2\Delta}{a_{\Delta}} \right)^n = \sqrt{1 + \xi},
\end{equation}

Define $a_{\max} = \max(a_0,a_{\Delta})$. According to \eqref{eq_sup_error_probability} and \eqref{eq_sup_density} there exists $a_* > a_{\max}$ s.t.
\begin{equation}
\frac{1}{M(S_{\mathbf{H}},a_*)} \sum_{s_{\text{rc}} \in S_{\mathbf{H}} \bigcap \mathbf{H}\cdot\text{Cb}(a_*)} P_e(s_{\text{rc}} | \mathbf{H}) \leq \sup_{b > a_{\max}} \frac{1}{M(S_{\mathbf{H}},b)} \sum_{s_{\text{rc}} \in S_{\mathbf{H}} \bigcap \mathbf{H}\cdot\text{Cb}(b)} P_e(s_{\text{rc}} | \mathbf{H}) < \epsilon(\mathbf{H})(1 + \xi/2)
\end{equation}
and
\begin{equation}
\label{eq_regularization_density_lower_bound}
\frac{M(S_{\mathbf{H}},a_*)}{\det(\mathbf{H})a_*^n} > \frac{\gamma_{\text{rc}}}{\sqrt{1 + \xi}}.
\end{equation}

Define the FC $G_{\mathbf{H}} = S_{\mathbf{H}} \bigcap \mathbf{H}\cdot\text{Cb}(a_*)$, and denote by $P_e^{G_{\mathbf{H}}}(s_{\text{rc}})$ the decoding error probability of any codeword $s_{\text{rc}} \in G_{\mathbf{H}}$ in transmission over the AWGN channel. Since $G_{\mathbf{H}} \subset S_{\mathbf{H}}$ then $P_e^{G_{\mathbf{H}}}(s_{\text{rc}}) \leq P_e(s_{\text{rc}}|\mathbf{H})$, and the average error probability of the FC is given by
\begin{equation}
\label{eq_FC_for_regularization_ub}
P_e(G_{\mathbf{H}}) = \frac{1}{|G_{\mathbf{H}}|} \sum_{s_{\text{rc}} \in G_{\mathbf{H}}} {P_e^{G_{\mathbf{H}}}(s_{\text{rc}})}
                 \leq \frac{1}{|G_{\mathbf{H}}|} \sum_{s_{\text{rc}} \in G_{\mathbf{H}}} {P_e(s_{\text{rc}}|\mathbf{H})}
                    < \epsilon(\mathbf{H})(1 + \xi/2).
\end{equation}

Now, we will create a regular IC, denoted by $S^{'}_{\mathbf{H}}$, by tiling the FC $G_{\mathbf{H}}$ to the whole space $\mathbb{R}^n$ in the
following way:
\begin{equation}
S^{'}_{\mathbf{H}} = \left\{ s_{\text{rc}} + \mathbf{H} \cdot I \cdot (a_* + 2\Delta) : s_{\text{rc}} \in G_{\mathbf{H}}, I \in \mathbb{Z}_n \right\},
\end{equation}
where $\mathbb{Z}_n$ is the $n$ dimensional integers lattice.

The error probability of any $s_{\text{rc}} \in S^{'}_{\mathbf{H}}$ equals the probability of decoding by a mistake to another codeword from the same copy of the FC $G_{\mathbf{H}}$ or to a codeword in another copy. Hence, the average error probability of $S^{'}_{\mathbf{H}}$, with equiprobable codewords transmission over the AWGN channel, can be upper bounded by the union bound as follows:
\begin{equation}
\label{eq_regularized_IC_upper_bound}
P_e\left(S^{'}_{\mathbf{H}}\right) \leq P_e\left(G_{\mathbf{H}}\right) + \sum_{i=1}^{n}{2Q\left(\frac{H_i\Delta}{\sigma}\right)}.
\end{equation}

Since the given fading channel realization is not a \emph{$\xi$ - strong fading realization}, and from the definition of $\Delta$ we obtain:
\begin{equation}
\label{eq_sum_of_Q_upper_bound}
\sum_{i=1}^{n}{2Q\left(\frac{H_i\Delta}{\sigma}\right)} \leq 2nQ\left(\frac{h^*_{\min}\Delta}{\sigma}\right) = \frac{\xi}{2}\cdot\epsilon(\mathbf{H}),
\end{equation}
where $h^*_{\min}(\xi)$ is the solution of $Pr\{H_{\min} \leq h^*_{\min} \} = \xi$. Combining \eqref{eq_FC_for_regularization_ub}, \eqref{eq_regularized_IC_upper_bound} and \eqref{eq_sum_of_Q_upper_bound} we obtain the desired result:
\begin{align}
P_e\left(S^{'}_{\mathbf{H}}\right) \leq \epsilon(\mathbf{H})(1 + \xi).
\end{align}

The density of $S^{'}_{\mathbf{H}}$ is given by
\begin{equation}
\gamma_{\text{rc}}^{'}(\mathbf{H}) = \gamma_{\text{rc}}^{'} = \frac{|G_{\mathbf{H}}|}{\text{Vol}(\mathbf{H}\cdot\text{Cb}(a_* + 2\Delta))}
                                                            = \frac{M(S_{\mathbf{H}}, a_*)}{\det(\mathbf{H})a_*^n}\cdot\left(\frac{a_*}{a_*+2\Delta} \right)^n.
\end{equation}

Combining \eqref{eq_regularization_density_lower_bound} with the definition of $a_{\Delta}$ and the fact that $a_* > a_{\Delta}$ we obtain the desired result:
\begin{align}
\gamma_{\text{rc}}^{'} > \frac{\gamma_{\text{rc}}}{1 + \xi}.
\end{align}

Let us denote by $\mathbf{H} = \text{diag}\{h_1, \dots, h_n\}$ the given channel realization. By its construction, for any $s_{\text{rc}} \in S^{'}_{\mathbf{H}}$, the set of points $\{s_{\text{rc}} \pm h_i\cdot (a_* + 2\Delta) \cdot \underline e_i, i=1, \dots, n \}$ is also in $S^{'}_{\mathbf{H}}$, where $\{\underline e_i\}_{i=1}^n$ is the standard basis of $\mathbb{R}^n$. Hence, any Voronoi cell of $S^{'}_{\mathbf{H}}$ is contained within a sphere of radius $r_0 \triangleq \sqrt{n}(a_* + 2\Delta)h_{\max}$ centered around its codeword, where $h_{\max} \triangleq \max(h_1, \dots, h_n)$. This proves that $S^{'}_{\mathbf{H}}$ is indeed a regular IC.
\end{IEEEproof}
\section{Proof of the Log of Chi Square Distribution Lemma}
\label{app_proof_lemma_ln_chi2_density_moments}
\begin{IEEEproof}[Proof of Lemma \ref{lem_ln_chi2_pdf}]
By simple variables substitution, we get the following relation between the CDFs of $Y_n$ and $X$:
\begin{equation}
\label{eq_Yn_Chi2_CDF}
F_{Y_n}(y) = F_{\chi^2_n}(ne^{\sqrt{\frac{2}{n}}y}).
\end{equation}
Then, if we differentiate \eqref{eq_Yn_Chi2_CDF} w.r.t. $y$ we will get the following relation between the RVs' PDFs:
\begin{equation}
\label{eq_Yn_Chi2_PDF}
f_{Y_n}(y) = \sqrt{2n}e^{\sqrt{\frac{2}{n}}y}f_{\chi^2_n}(ne^{\sqrt{\frac{2}{n}}y}).
\end{equation}
Assignment of the $\chi^2_n$'s PDF, $f_{\chi^2_n}(x) = \frac{x^{\frac{n}{2}-1}e^{-\frac{x}{2}}}{2^\frac{n}{2}\Gamma(\frac{n}{2})}, x>0$ will give us
\begin{equation}
f_{Y_n}(y) = \frac{ (\frac{n}{2})^{\frac{n-1}{2}}}{\Gamma(\frac{n}{2})}e^{\sqrt{\frac{n}{2}}y - \frac{n}{2}e^{\sqrt{\frac{2}{n}}y}} \nonumber,
\end{equation}
which completes the proof of \eqref{eq_Yn_PDF}.
From the Stirling approximation for the Gamma function for $z\in\mathbb{R}$ we get
\begin{equation}
\label{eq_Gamma_Stirling_approximation}
\Gamma(z+1)=z\Gamma(z)=\sqrt{2\pi e}\left(\frac{z}{e}\right)^z\left(1+O\left(\frac{1}{z}\right)\right).
\end{equation}
Using \eqref{eq_Gamma_Stirling_approximation} for $z=\frac{n}{2}$ we get
\begin{equation}
\label{eq_Gamma__n_div_2_approximation}
\Gamma\left(\frac{n}{2}\right)=\frac{\Gamma(\frac{n}{2}+1)}{\frac{n}{2}}=\sqrt{\frac{4\pi}{n}}\left(\frac{n}{2e}\right)^{\frac{n}{2}}\left(1+O\left(\frac{1}{n}\right)\right).
\end{equation}
The assignment of \eqref{eq_Gamma__n_div_2_approximation} in \eqref{eq_Yn_PDF} gives us
\begin{align}
\begin{aligned}
\label{align_Yn_PDF_approximation_1}
f_{Y_n}(y) &= \frac{1}{\sqrt{2\pi}}e^{\frac{n}{2} + \sqrt{\frac{n}{2}}y - \frac{n}{2}e^{\sqrt{\frac{2}{n}}y}}\left(\frac{1}{1+O\left(\frac{1}{n}\right)}\right)
\\         &= \frac{1}{\sqrt{2\pi}}e^{\frac{n}{2} + \sqrt{\frac{n}{2}}y - \frac{n}{2}e^{\sqrt{\frac{2}{n}}y}}\left(1+O\left(\frac{1}{n}\right)\right),
\end{aligned}
\end{align}
for any $n > N_0$, for some finite $N_0$.
\newline\indent
By Taylor's theorem for $g(x) = e^x$ around $x_0 = 0$, the following holds:
\begin{equation}
g(x) = \sum_{k=0}^{K}\frac{x^k}{k!} + \frac{e^{\zeta}x^{K+1}}{(K+1)!},
\end{equation}
for some real number $\zeta \in [0,x]$.
Using it with $K=2$ and $x \equiv \sqrt{\frac{2}{n}}y$ we obtain:
\begin{equation}
e^{\sqrt{\frac{2}{n}}y} = 1 + \sqrt{\frac{2}{n}}y + {\frac{1}{n}}y^2 + \frac{\sqrt{2}e^{\zeta(y)}}{3n^{\frac{3}{2}}}y^3,
\end{equation}
where for $y\in[-n^{\frac{1}{6}},n^{\frac{1}{6}}]$, then $\zeta(y)\in[-\frac{\sqrt{2}}{n^{\frac{1}{3}}},\frac{\sqrt{2}}{n^{\frac{1}{3}}}]$.
\newline
Assigning it in \eqref{align_Yn_PDF_approximation_1} , for any $n > N_0$ and for $y\in[-n^{\frac{1}{6}},n^{\frac{1}{6}}]$, gives us:
\begin{equation}
\label{eq_Yn_PDF_approximation_2}
f_{Y_n}(y) = \frac{1}{\sqrt{2\pi}}e^{-\frac{y^2}{2}} \cdot e^{-\frac{e^{\zeta(y)}}{3\sqrt{2n}}y^3}\left({1+O\left(\frac{1}{n}\right)}\right).
\end{equation}
Using Taylor's theorem again with $K=0$ and $x \equiv -\frac{e^{\zeta(y)}}{3\sqrt{2n}}y^3$ we obtain:
\begin{equation}
e^{-\frac{e^{\zeta(y)}}{3\sqrt{2n}}y^3} = 1 - \frac{e^{-\eta(y)} \cdot e^{\zeta(y)}}{3\sqrt{2n}}y^3,
\end{equation}
where for $y\in[-n^{\frac{1}{6}-\delta},n^{\frac{1}{6}-\delta}]$ for some $0 \leq \delta < \frac{1}{6}$, then $\eta(y)\in(-\frac{1}{n^{3\delta}},\frac{1}{n^{3\delta}})$.
\newline
Combining all the above, we get that for any $n > N_0$, and for $y\in[-n^{\frac{1}{6}-\delta},n^{\frac{1}{6}-\delta}]$ for some $0 \leq \delta < \frac{1}{6}$:
\begin{equation}
\label{eq_Yn_PDF_approximation_2}
f_{Y_n}(y) = \frac{1}{\sqrt{2\pi}}e^{-\frac{y^2}{2}} - \frac{e^{\nu(y)}}{6\sqrt{\pi}}\cdot\frac{y^3e^{-\frac{y^2}{2}}}{\sqrt{n}} + O\left(\frac{e^{-\frac{y^2}{2}}}{n}\right),
\end{equation}
where $\nu(y)\triangleq\zeta(y)-\eta(y)$ and $|\nu(y)| < \frac{1}{n^{3\delta}} + \frac{\sqrt{2}}{n^{\frac{1}{3}}}$.

By definition $e_n(y) \triangleq f_{Y_n}(y) - N(0,1)$, then:
\begin{align}
\begin{aligned}
e_n &\triangleq \int_{-\infty}^{\infty}|e_n(y)|dy
\\  &\leq \int_{|y|\leq n^{\frac{1}{6}}}|e_n(y)|dy + \int_{|y|>n^{\frac{1}{6}}}f_{Y_n}(y)dy + \int_{|y|>n^{\frac{1}{6}}}N(0,1)dy
\\  &= \int_{|y|\leq n^{\frac{1}{6}}}|e_n(y)|dy + 1 - \int_{|y|\leq n^{\frac{1}{6}}}f_{Y_n}(y)dy + \int_{|y|>n^{\frac{1}{6}}}N(0,1)dy
\\  &= \int_{|y|\leq n^{\frac{1}{6}}}|e_n(y)|dy + 1 - \int_{|y|\leq n^{\frac{1}{6}}}N(0,1)dy - \int_{|y|\leq n^{\frac{1}{6}}}e_n(y)dy + \int_{|y|>n^{\frac{1}{6}}}N(0,1)dy
\\  &= \int_{|y|\leq n^{\frac{1}{6}}}|e_n(y)|dy - \int_{|y|\leq n^{\frac{1}{6}}}e_n(y)dy + 2\int_{|y|>n^{\frac{1}{6}}}N(0,1)dy
\\  &\leq 2\int_{|y|\leq n^{\frac{1}{6}}}|e_n(y)|dy + 4Q(n^{\frac{1}{6}})
\\  &= O\left(\int_{-\infty}^{\infty}\frac{|y|^3e^{-\frac{y^2}{2}}}{\sqrt{n}}dy\right) + O\left(\int_{-\infty}^{\infty}\frac{e^{-\frac{y^2}{2}}}{n}dy\right) + O\left(e^{-\frac{n^{\frac{1}{3}}}{2}}\right) = O\left(\frac{1}{\sqrt{n}}\right),
\end{aligned}
\end{align}
which completes the proof of \eqref{eq_Yn_asymptotic_PDF}.
\end{IEEEproof}
\section{Proof of the Sum of Two Almost Normal RVs Lemma}
\label{app_proof_lemma_sum_of_almost_normal_RVs}
\begin{IEEEproof}[Proof of Lemma \ref{lem_sum_of_almost_normal_RVs}]
$X_1$ and $X_2$ are independent. Hence, by definition, the CDF of $Y$ is given by
\begin{align}
\begin{aligned}
F_Y(y) &\triangleq Pr\{ Y \leq y \}
\\     &= Pr\{ X_1 + X_2 \leq y \}
\\     &= \int_{-\infty}^{y}{f_{X_1}(x) \cdot F_{X_2}(y - x)}dx.
\end{aligned}
\end{align}
By the assignment of $f_{X_1}(x)$ and $F_{X_2}(x)$ given by the lemma, we can obtain the following:
\begin{align}
\begin{aligned}
F_Y(y) &= \int_{-\infty}^{y}{N(0,\sigma_1^2) \cdot F_{N(0,\sigma_2^2)}(y - x)}dx
\\     &+ O\left( \int_{-\infty}^{y}{ e_n(x) \cdot F_{N(0,\sigma_2^2)}(y - x)}dx \right)
\\     &+ O\left( \int_{-\infty}^{y}{ \frac{N(0,\sigma_1^2)}{\sqrt{n}}}dx \right)
\\     &+ O\left( \int_{-\infty}^{y}{ \frac{e_n(x)}{\sqrt{n}} }dx \right).
\end{aligned}
\end{align}
Since, $F_{N(0,\sigma_y^2)}(y) = \int_{-\infty}^{y}{N(0,\sigma_1^2) \cdot F_{N(0,\sigma_2^2)}(y - x)}dx$, we can get
\begin{align}
\begin{aligned}
|F_Y(y) - F_{N(0,\sigma_y^2)}(y)| &\leq O\left( \int_{-\infty}^{\infty}{ |e_n(x)| }dx \right)
+ O\left( \int_{-\infty}^{\infty}{ \frac{N(0,\sigma_1^2)}{\sqrt{n}}}dx \right)
+ O\left( \int_{-\infty}^{\infty}{ \frac{|e_n(x)|}{\sqrt{n}} }dx \right)
\\ &= O\left(\frac{1}{\sqrt{n}}\right) + O\left(\frac{1}{\sqrt{n}}\right) + O\left(\frac{1}{n}\right) = O\left(\frac{1}{\sqrt{n}}\right),
\end{aligned}
\end{align}
which completes the proof of \eqref{eq_sum_of_almost_normal_CDF}.
\end{IEEEproof}
\section{Lemma 6}
\label{app_lemma_47_in_Polyanskiy}
\begin{lem}
\label{lem_47_in_Polyanskiy}
Let $Z_1,Z_2,\dots,Z_n$ be independent random variables, $\sigma^2=\sum_{i=1}^{n}{Var(Z_i)}$ be non-zero and $T=\sum_{i=1}^{n}{E\{|Z_i-E\{Z_i\}|^3\}}<\infty$; then for any $A$
\begin{equation}
E\left\{e^{-\sum_{i=1}^{n}{Z_i}}1_{\left\{\sum_{i=1}^{n}{Z_i}>A\right\}}\right\}  \leq 2\left(\frac{\ln(2)}{\sqrt{2\pi}} + \frac{12T}{\sigma^2}\right)\frac{1}{\sigma}e^{-A}.
\end{equation}
\end{lem}
\begin{IEEEproof}
See Lemma 47 in \cite{Polyanskiy}.
\end{IEEEproof}
\section{The Channel Output Given CSI Distribution Lemma}
\label{app_lemma_Y_given_H_pdf}
\begin{lem}
\label{lem_Y_given_H_PDF}
Suppose that $Y = H\cdot X + Z$, where $X\sim U(-\frac{a}{2},\frac{a}{2})$ and $Z\sim N(0, \sigma^2)$ are independent. If $H$ is also a random variable independent of $X$ and $Z$, then
\begin{equation}
\label{eq_Y_given_H_PDF}
f(y|h) = \frac{1}{ah}\left(Q\left(\frac{y}{\sigma} - \frac{ah}{2\sigma}\right) - Q\left(\frac{y}{\sigma} + \frac{ah}{2\sigma}\right)\right).
\end{equation}
\end{lem}
\begin{IEEEproof}
\begin{align}
\label{align_Y_given_H_PDF_prrof}
\begin{aligned}
f(y|h) &= \int_{-\infty}^{\infty}{f(y,x|h)}dx
\\     &= \int_{-\infty}^{\infty}{f(x)f(y|x,h)}dx
\\     &= \int_{-\frac{a}{2}}^{\frac{a}{2}}{\frac{1}{a}\cdot\frac{1}{\sqrt{2\pi\sigma^2}}e^{-\frac{(y-hx)^2}{2\sigma^2}}}dx
\\     &= \int_{-\frac{ah}{2}}^{\frac{ah}{2}}{\frac{1}{ah}\cdot\frac{1}{\sqrt{2\pi\sigma^2}}e^{-\frac{(x-y)^2}{2\sigma^2}}}dx
\\     &= \frac{1}{ah}\left( \int_{-\frac{ah}{2}}^{\infty}{\frac{1}{\sqrt{2\pi\sigma^2}}e^{-\frac{(x-y)^2}{2\sigma^2}}}dx - \int_{\frac{ah}{2}}^{\infty}{\frac{1}{\sqrt{2\pi\sigma^2}}e^{-\frac{(x-y)^2}{2\sigma^2}}}dx \right)
\\     &= \frac{1}{ah}\left( Q\left(-\frac{ah}{2\sigma} - \frac{y}{\sigma}\right) - Q\left(\frac{ah}{2\sigma} - \frac{y}{\sigma}\right) \right)
\\     &= \frac{1}{ah}\left( Q\left(\frac{y}{\sigma} - \frac{ah}{2\sigma}\right) - Q\left(\frac{y}{\sigma} + \frac{ah}{2\sigma}\right)\right).
\end{aligned}
\end{align}
\end{IEEEproof}
\section{Proof of the Information Density's Moments Lemma}
\label{app_proof_lemma_information_density_moments}
\begin{IEEEproof}[Proof of Lemma \ref{lem_information_density_moments}]
\subsection{Preliminary Analysis}
The information density is given by
\begin{align}
\begin{aligned}
i(x;y,h) &\triangleq \ln\left(\frac{f(x,y,h)}{f(x)f(y,h)}\right)
\\       &= \ln\left(\frac{f(x)f(h)f(y|h,x)}{f(x)f(h)f(y|h)}\right)
\\       &= \ln\left(\frac{f(y|h,x)}{f(y|h)}\right)
\\       &= \ln\left(\frac{f(z=y-hx)}{f(y|h)}\right)
\\       &= \ln\left( \frac{1}{\sqrt{2\pi\sigma^2}}e^{-\frac{z^2}{2\sigma^2}} \right) - \ln\left(\frac{1}{ah}\left(Q\left(\frac{y}{\sigma} - \frac{ah}{2\sigma}\right) - Q\left(\frac{y}{\sigma} + \frac{ah}{2\sigma}\right)\right)\right)
\\       &= \frac{1}{2}\ln\left(\frac{a^2h^2}{2\pi e\sigma^2}\right) - \frac{z^2-\sigma^2}{2\sigma^2} - \ln\left(Q\left(\frac{y}{\sigma} - \frac{ah}{2\sigma}\right) - Q\left(\frac{y}{\sigma} + \frac{ah}{2\sigma}\right)\right)
\\       &= \frac{1}{2}\ln\left(\frac{a^2h^2}{2\pi e\sigma^2}\right) - \frac{z^2-\sigma^2}{2\sigma^2} + e_{a/\sigma}(y,h)
\end{aligned}
\end{align}
where $f(y|h)$ is given by Lemma \ref{lem_Y_given_H_PDF} (see Appendix \ref{app_lemma_Y_given_H_pdf}) and the following definition of
\begin{equation}
e_{a/\sigma}(y,h) \triangleq - \ln\left(Q\left(\frac{y}{\sigma} - \frac{ah}{2\sigma}\right) - Q\left(\frac{y}{\sigma} + \frac{ah}{2\sigma}\right)\right)\geq0.
\end{equation}
Define the three error's moments for $i=1,2,3$ by
\begin{align}
e_{a/\sigma,i} &\triangleq E\{e_{a/\sigma}^{i}(Y,H)\}
\\             &= E\{E\{e_{a/\sigma}^{i}(Y,H)|H\}\}
\\             &= E\{e_{a/\sigma,i}(H)\}
\end{align}
where
\begin{align}
\begin{aligned}
e_{a/\sigma,i}(h) &\triangleq E\{e_{a/\sigma}^{i}(Y,H)|H=h\}
\\              &= (-1)^i\int_{-\infty}^{\infty}{\frac{1}{ah}\left(Q\left(\frac{y}{\sigma} - \frac{ah}{2\sigma}\right) - Q\left(\frac{y}{\sigma} + \frac{ah}{2\sigma}\right)\right)\ln^{i}\left(Q\left(\frac{y}{\sigma} - \frac{ah}{2\sigma}\right) - Q\left(\frac{y}{\sigma} + \frac{ah}{2\sigma}\right)\right)}dy
\\              &= (-1)^i\frac{\sigma}{ah}\int_{-\infty}^{\infty}{\left(Q\left(y - \frac{ah}{2\sigma}\right) - Q\left(y + \frac{ah}{2\sigma}\right)\right)\ln^i\left(Q\left(y - \frac{ah}{2\sigma}\right) - Q\left(y + \frac{ah}{2\sigma}\right)\right)}dy
\\              &= \frac{\sigma}{ah}\eta_i\left(\frac{ah}{\sigma}\right)
\end{aligned}
\end{align}
and
\begin{equation}
\eta_i\left(\frac{ah}{\sigma}\right) \triangleq (-1)^{i}\int_{-\infty}^{\infty}{\left(Q\left(y - \frac{ah}{2\sigma}\right) - Q\left(y + \frac{ah}{2\sigma}\right)\right)\ln^{i}\left(Q\left(y - \frac{ah}{2\sigma}\right) - Q\left(y + \frac{ah}{2\sigma}\right)\right)}dy, i = 1,2,3.
\end{equation}
As can be seen in Fig. \ref{fig_eta_Vs_h}, the function $\eta_i\left(\frac{ah}{\sigma}\right)$ is nonnegative, bounded and asymptotically converges to a constant for $h\in(0,\infty)$ and for $i = 1,2,3$. The function is also monotonically nondecreasing for $i=1$.
For small values of $ah/\sigma$, we can approximate $\eta_i\left(\frac{ah}{\sigma}\right)$ by
\begin{align}
\begin{aligned}
\eta_i\left(\frac{ah}{\sigma}\right) &= -(-1)^{i}\frac{ah}{\sigma}\int_{-\infty}^{\infty}{\frac{Q(y + \frac{ah}{2\sigma}) - Q(y - \frac{ah}{2\sigma})}{\frac{ah}{\sigma}}\ln^{i}\left(-\frac{ah}{\sigma}\frac{Q(y + \frac{ah}{2\sigma}) - Q(y - \frac{ah}{2\sigma})}{\frac{ah}{\sigma}}\right)}dy
\\                        &\approx -(-1)^{i}\frac{ah}{\sigma}\int_{-\infty}^{\infty}{Q'(y)\ln^{i}\left(-\frac{ah}{\sigma}Q'(y)\right)}dy
\\                        &= (-1)^{i}\frac{ah}{\sigma}\int_{-\infty}^{\infty}{\frac{1}{\sqrt{2\pi}}e^{-\frac{y^2}{2}}\ln^{i}\left(\frac{ah}{\sigma}\frac{1}{\sqrt{2\pi}}e^{-\frac{y^2}{2}}\right)}dy
\\                        &= (-1)^{i}\frac{ah}{\sigma}\int_{-\infty}^{\infty}{\frac{1}{\sqrt{2\pi}}e^{-\frac{y^2}{2}}\left(C(ah/\sigma) + \frac{1-y^2}{2}\right)^{i}}dy
\\                        &= (-1)^{i}\frac{ah}{\sigma}E_{N(0,1)}\left\{ \left(C(ah/\sigma) + \frac{1-y^2}{2}\right)^{i} \right\}
\end{aligned}
\end{align}
where
\begin{equation}
C(ah/\sigma) \triangleq \frac{1}{2}\ln\left(\frac{a^2h^2}{2\pi e\sigma^2 }\right).
\end{equation}
By simple calculation of the moments of a standard normal random variable, we get that for small values of $ah/\sigma$
\begin{align}
\begin{aligned}
\eta_1\left(\frac{ah}{\sigma}\right) &\approx -\frac{ah}{\sigma}C(ah/\sigma),
\\
\eta_2\left(\frac{ah}{\sigma}\right) &\approx \frac{ah}{\sigma}\left(C(ah/\sigma)^2 + \frac{1}{2}\right),
\\
\eta_3\left(\frac{ah}{\sigma}\right) &\approx -\frac{ah}{\sigma}\left(C(ah/\sigma)^3 + \frac{3}{2}C(ah/\sigma) - 1\right).
\end{aligned}
\end{align}
It can be seen in Fig. \ref{fig_eta_Vs_h} that for $ah/\sigma < 1$ the approximations above are very accurate.

First, let us calculate the first order error's moment
\begin{align}
\label{align_ea_definition}
e_{a/\sigma,1} &\triangleq E\left\{e_{a/\sigma}(H)\right\}
\\           &= E\left\{\frac{\sigma}{aH}\eta_1\left(\frac{aH}{\sigma}\right)\right\}
\\           &= \int_{0}^{\infty}{f(h)\frac{\sigma}{ah}\eta_1\left(\frac{ah}{\sigma}\right)}dh
\\           &= \int_{0}^{\frac{\sigma}{a}}{f(h)\frac{\sigma}{ah}\eta_1\left(\frac{ah}{\sigma}\right)}dh + \int_{\frac{\sigma}{a}}^{\infty}{f(h)\frac{\sigma}{ah}\eta_1\left(\frac{ah}{\sigma}\right)}dh.
\label{align_ea_as_sum_of_integrals}
\end{align}
For any regular fading distribution there exists a positive constant $\beta > 0$ s.t. near the origin $f(h)\sim\frac{1}{h^{1-\beta}}$. Moreover, for any PDF there exists a positive constant $\beta^{'} > 0$ s.t. $f(h)\sim\frac{1}{h^{1+\beta^{'}}}$ for large enough $h$. Hence, for large enough $a/\sigma$, we can get the following bounds
\begin{enumerate}
  \item
    \begin{align}
    \int_{0}^{\frac{\sigma}{a}}{f(h)\frac{\sigma}{ah}\eta_1\left(\frac{ah}{\sigma}\right)}dh &= O\left(-\int_{0}^{\frac{\sigma}{a}}{f(h)C(ah/\sigma)}dh\right)
    \\ &= O\left(\int_{0}^{\frac{\sigma}{a}}{\frac{1}{h^{1-\beta}}\ln\left(\frac{ah}{\sigma}\right)}dh\right)
    \\ &= O\left(\int_{0}^{\frac{\sigma}{a}}{\frac{\ln(h)}{h^{1-\beta}}}dh\right) + O\left(\ln\left(\frac{a}{\sigma}\right)\int_{0}^{\frac{\sigma}{a}}{\frac{dh}{h^{1-\beta}}}\right)
    \\ \label{align_bounding_ea_part1}
       &= O\left(\ln\left(\frac{a}{\sigma}\right)\left(\frac{\sigma}{a}\right)^{\beta}\right).
    \end{align}
  \item
    \begin{align}
    \label{align_bounding_eta1}
    \int_{\frac{\sigma}{a}}^{\infty}{f(h)\frac{\sigma}{ah}\eta_1\left(\frac{ah}{\sigma}\right)}dh &\leq  \int_{\frac{\sigma}{a}}^{\infty}{f(h)\frac{\sigma}{ah}M}dh
    \\ &= \int_{\frac{\sigma}{a}}^{h_0}{f(h)\frac{\sigma}{ah}M}dh
        + \int_{h_0}^{h_1}{f(h)\frac{\sigma}{ah}M}dh
        + \int_{h_1}^{\infty}{f(h)\frac{\sigma}{ah}M}dh
    \\ &= O\left(\frac{\sigma}{a}\int_{\frac{\sigma}{a}}^{\infty}{\frac{dh}{h^{2-\beta}}}\right)
        + O\left(\frac{\sigma}{a}\int_{h_0}^{h_1}{\frac{f(h)}{h}dh}\right)
        + O\left(\frac{\sigma}{a}\int_{h_1}^{\infty}{\frac{dh}{h^{2+\beta^{'}}}}\right)
    \\ \label{align_bounding_ea_part2}
       &= O\left(\left(\frac{\sigma}{a}\right)^{\beta}\right)
        + O\left(\frac{\sigma}{a}\right)
        = O\left(\left(\frac{\sigma}{a}\right)^{\min(\beta,1)}\right),
    \end{align}
\end{enumerate}
where \eqref{align_bounding_eta1} is due to the fact that $\eta_1(ah/\sigma)\leq M$ for some positive and finite constant $M$.
From \eqref{align_ea_as_sum_of_integrals}, \eqref{align_bounding_ea_part1}, \eqref{align_bounding_ea_part2} and the fact that $\forall\epsilon>0~\lim_{x\to\infty}\frac{\ln(x)}{x^{\epsilon}}=0$, we get that there exists a constant $0 < \alpha \leq 1$ s.t. the following holds:
\begin{equation}
e_{a/\sigma,1} = O\left(\left(\frac{\sigma}{a}\right)^{\alpha}\right).
\end{equation}
Finally, because of the common properties of $\eta_1(ah/\sigma), \eta_2(ah/\sigma)$ and $\eta_3(ah/\sigma)$, with equivalent calculations, we can get that there exists also a constant $0 < \alpha \leq 1$, s.t the error's moments hold the following:
\begin{equation}
e_{a/\sigma,2} \triangleq E\left\{e_{a/\sigma}^{2}(Y,H)\right\}= E\left\{\frac{\sigma}{aH}\eta_2\left(\frac{aH}{\sigma}\right)\right\} = O\left(\left(\frac{\sigma}{a}\right)^{\alpha}\right),
\end{equation}
\begin{equation}
e_{a/\sigma,3} \triangleq E\left\{|e_{a/\sigma}(Y,H)|^{3}\right\} = E\left\{\frac{\sigma}{aH}\eta_3\left(\frac{aH}{\sigma}\right)\right\} = O\left(\left(\frac{\sigma}{a}\right)^{\alpha}\right).
\end{equation}
\begin{figure}[htp]
\psfrag{ah/sigma}{$\nicefrac{ah}{\sigma}$}
\psfrag{etai}{$\eta_i$}
\psfrag{eta1}{$\eta_1$}
\psfrag{eta2}{$\eta_2$}
\psfrag{eta3}{$\eta_3$}
\psfrag{eta1 approx.}{$\eta_1~ approx.$}
\psfrag{eta2 approx.}{$\eta_2~ approx.$}
\psfrag{eta3 approx.}{$\eta_3~ approx.$}
\center{\includegraphics[width=1\columnwidth]{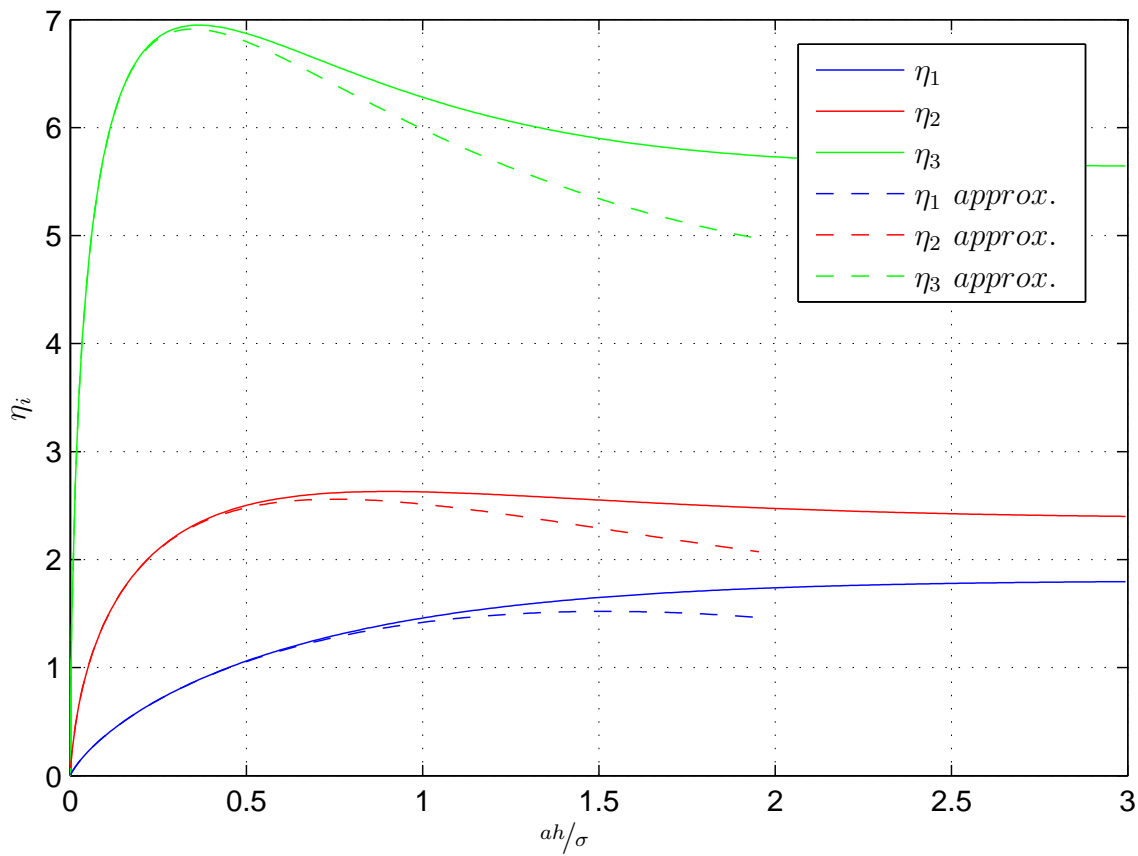}}
\caption{\label{fig_eta_Vs_h} $\eta_{i}(ah/\sigma)$ and its approximation for small values of $ah/\sigma$.}
\end{figure}
\subsection{Calculating the Mutual Information}
The mean of the information density is given by
\begin{align}
I(X;Y,H) &\triangleq E\{i(X;Y,H)\}
\\       &= E\left\{\frac{1}{2}\ln\left(\frac{a^2H^2}{2\pi e\sigma^2}\right)\right\} - E\left\{\frac{Z^2-\sigma^2}{2\sigma^2}\right\} + E\left\{e_{a/\sigma}(Y,H)\right\}
\\       &= E\left\{\frac{1}{2}\ln\left(\frac{a^2H^2}{2\pi e\sigma^2}\right)\right\} + e_{a/\sigma,1}
\\       &= E\left\{\frac{1}{2}\ln\left(\frac{a^2H^2}{2\pi e\sigma^2}\right)\right\} + O\left(\left(\frac{\sigma}{a}\right)^{\alpha}\right).
\end{align}
\subsection{Calculating the Information Density Variance}
The variance of the information density is given by
\begin{align}
Var(i(X;Y,H)) &= Var\left(\frac{1}{2}\ln\left(\frac{a^2H^2}{2\pi e\sigma^2}\right) - \frac{Z^2-\sigma^2}{2\sigma^2} + e_{a/\sigma}(Y,H)\right)
\\            &= Var\left(\frac{1}{2}\ln\left(H^2\right) - \frac{Z^2}{2\sigma^2} + e_{a/\sigma}(Y,H)\right)
\\            &= Var\left(\frac{1}{2}\ln\left(H^2\right)\right) + Var\left(\frac{Z^2}{2\sigma^2}\right) + Var\left(e_{a/\sigma}(Y,H)\right)
\\            &+ 2Cov\left(\frac{1}{2}\ln\left(H^2\right),e_{a/\sigma}(Y,H)\right) -2Cov\left(\frac{Z^2}{2\sigma^2},e_{a/\sigma}(Y,H)\right)
\\            \label{align_var_calc}
              &= \frac{1}{2} + Var\left(\delta(H)\right) + \Delta(a/\sigma)
\end{align}
where
\begin{align}
\begin{aligned}
\label{align_line_in_var_bound}
\Delta(a/\sigma) &\triangleq Var\left(e_{a/\sigma}(Y,H)\right) + 2Cov\left(\frac{1}{2}\ln\left(H^2\right),e_{a/\sigma}(Y,H)\right) -2Cov\left(\frac{Z^2}{2\sigma^2},e_{a/\sigma}(Y,H)\right)
\\ &= e_{a/\sigma,2} - e_{a/\sigma,1}^{2} + E\left\{\ln\left(H^2\right)e_{a/\sigma}(Y,H)\right\} - E\left\{\ln\left(H^2\right)\right\}e_{a/\sigma,1}
\\ &- E\left\{\frac{Z^2}{\sigma^2}e_{a/\sigma}(Y,H)\right\} + E\left\{\frac{Z^2}{\sigma^2}\right\}e_{a/\sigma,1}
\\ &= O(e_{a/\sigma,2}) + O(e_{a/\sigma,1}) + O\left(E\left\{\ln\left(H^2\right)e_{a/\sigma}(Y,H)\right\}\right) + O\left(E\left\{\frac{Z^2}{\sigma^2}e_{a/\sigma}(y,h)\right\}\right).
\end{aligned}
\end{align}
By the Cauchy Schwarz inequality,
\begin{equation}
\label{eq_cauchy_schwarz_1}
\left|E\left\{\ln\left(H^2\right)e_{a/\sigma}(Y,H)\right\}\right| \leq \sqrt{E\left\{\ln^{2}\left(H^2\right)\right\}e_{a/\sigma,2}} = O\left(\sqrt{e_{a/\sigma,2}}\right)
\end{equation}
and
\begin{equation}
\label{eq_cauchy_schwarz_2}
\left|E\left\{\frac{Z^2}{\sigma^2}e_{a/\sigma}(Y,H)\right\}\right| \leq \sqrt{E\left\{\frac{Z^4}{\sigma^4}\right\}e_{a/\sigma,2}} = O\left(\sqrt{e_{a/\sigma,2}}\right).
\end{equation}
Combining \eqref{align_line_in_var_bound}, \eqref{eq_cauchy_schwarz_1} and \eqref{eq_cauchy_schwarz_2} we get
\begin{equation}
\label{eq_var_error_bound}
\Delta(a/\sigma) = O\left(\sqrt{e_{a/\sigma,2}}\right) = O\left(\left(\frac{\sigma}{a}\right)^{\frac{\alpha}{2}}\right).
\end{equation}
From \eqref{align_var_calc} and \eqref{eq_var_error_bound} we get the desired result:
\begin{equation}
Var(i(X;Y,H)) = \frac{1}{2} + Var\left(\delta(H)\right) + O\left(\left(\frac{\sigma}{a}\right)^{\frac{\alpha}{2}}\right).
\end{equation}
\subsection{Bounding the Information Density's Absolute third Order Moment}
The absolute third order moment of the information density is given by
\begin{align}
\rho_3 &\triangleq E\left\{|i(X;Y,H) - I(X;Y,H)|^3\right\}
\\     &= E\left\{\left|\frac{1}{2}\ln\left(\frac{a^2H^2}{2\pi e\sigma^2}\right) - \frac{Z^2-\sigma^2}{2\sigma^2} + e_{a/\sigma}(Y,H) - E\left\{\frac{1}{2}\ln\left(\frac{a^2H^2}{2\pi e\sigma^2}\right)\right\} - e_{a/\sigma,1}\right|^3\right\}
\\     \label{align_bounding_abs_3_moment}
       &\leq \left( \Big\| \frac{1}{2}\ln\left(H^2\right) - E\left\{\frac{1}{2}\ln\left(H^2\right)\right\} \Big\|_3 + \Big\| \frac{Z^2-\sigma^2}{2\sigma^2} \Big\|_3 + \Big\| e_{a/\sigma}(Y,H)\Big\|_3 + e_{a/\sigma,1} \right)^3
\end{align}
where the last inequality is due to the Minkowski inequality and the definition of $\left\|X\right\|_3 \triangleq E\left\{ |X|^3 \right\}^{\frac{1}{3}}$.
By definition we get
\begin{equation}
\label{eq_norm_3_of_ea_3}
\Big\| e_{a/\sigma}(Y,H)\Big\|_3 = \left(E\left\{e_{a/\sigma}^{3}(Y,H)\right\}\right)^{\frac{1}{3}} = e_{a/\sigma,3}^{\frac{1}{3}} = O\left(\ln\left(\frac{a}{\sigma}\right)\left(\frac{\sigma}{a}\right)^{\frac{\alpha}{3}}\right).
\end{equation}
From \eqref{align_bounding_abs_3_moment} and \eqref{eq_norm_3_of_ea_3} we get the desired result
\begin{equation}
\rho_3 \leq A + O\left(\ln\left(\frac{a}{\sigma}\right)\left(\frac{\sigma}{a}\right)^{\frac{\alpha}{3}}\right)
\end{equation}
for some positive and finite constant $A$, or simply $\rho_3 < \infty$.
\end{IEEEproof}
\section{Tiling}
\label{app_tiling}
We now turn to construct an IC with average error probability which is upper bounded by $\epsilon$, denoted by $S(n,\epsilon)$, from the FC  $S(n,\epsilon',a/\sigma)$. It is assumed that $S(n,\epsilon',a/\sigma)$ has an average error probability which is upper bounded by $\epsilon'$ (using the suboptimal decoder on which the dependence testing bound is based), and its NLD, $\delta(n,\epsilon',a/\sigma)$ in $\text{Cb}(a)$, holds the following:
\begin{equation}
\delta(n,\epsilon',a/\sigma) = \delta^* - \sqrt{ \frac{V}{n} }Q^{-1}(\epsilon') +
O\left(\frac{1}{n} + \frac{1}{\sqrt{n}}\left(\frac{\sigma}{a}\right)^{\frac{\alpha}{2}}  + \left(\frac{\sigma}{a}\right)^{\alpha}\right).
\end{equation}

Define the IC $S(n,\epsilon)$ as an infinite replication of $S(n,\epsilon',a/\sigma)$ with spacing of $b$ between every two copies as follows:
\begin{equation}
S(n,\epsilon) \triangleq \left\{ s + I\cdot(a+b): s\in S(n,\epsilon',a/\sigma), I\in \mathbb{Z}_n \right\}
\end{equation}
where $\mathbb{Z}_n$ denotes the integer lattice of dimension $n$.
The NLD of the IC is given by
\begin{align}
\begin{aligned}
\label{align_IC_NLD}
\delta(n,\epsilon,a/\sigma,b) &\triangleq \frac{1}{n}\ln\left(\frac{M(n,\epsilon',a/\sigma)}{(a+b)^n}\right)
\\                            &= \frac{1}{n}\ln\left(\frac{M(n,\epsilon',a/\sigma)}{a^n}\right) - \ln\left(1+\frac{b}{a}\right)
\\                            &= \delta(n,\epsilon',a/\sigma) - \ln\left(1+\frac{b}{a}\right),
\end{aligned}
\end{align}
where $M(n,\epsilon',a/\sigma)$ is the number of codewords of the FC.

Define the faded FC in the receiver, given the CSI, as
\begin{equation}
S(n,\epsilon',a/\sigma)_{\mathbf{H}} \triangleq \{ \mathbf{H}\cdot s: s\in S(n,\epsilon',a/\sigma) \}
\end{equation}
where $\mathbf{H}=\text{diag}\{H_1,H_2,\dots,H_n\}$.
In the receiver, we get the following IC:
\begin{equation}
S(n,\epsilon)_{\mathbf{H}} \triangleq \left\{ s_{\text{rc}} + \mathbf{H} \cdot I\cdot (a+b): s_{\text{rc}}\in S(n,\epsilon',a/\sigma)_{\mathbf{H}}, I\in \mathbb{Z}_n \right\},
\end{equation}
which is a tiled version of the faded FC.

Now consider the ML error probability of a point $s_{\text{rc}}\in S(n,\epsilon)_{\mathbf{H}}$, given the CSI $\mathbf{H}$ at the receiver, denoted by $P_{e,ML}^{IC}(s_{\text{rc}}|\mathbf{H})$. In the same manner, $P_{e,ML}^{FC}(s_{\text{rc}}|\mathbf{H})$ will denote the ML error probability for any $s_{\text{rc}}\in S(n,\epsilon',a/\sigma)_{\mathbf{H}}$.
If $\mathbf{H}$ is a too ``strong'' channel fading realization then we will declare an error. Formally, if $H_{\min} \leq h_{\min}^{*}$ for some arbitrary positive constant $h_{\min}^{*}$, where $H_{\min} \triangleq \min\{H_1,H_2,\dots,H_n\}$, then we will declare an error. Otherwise, this error probability equals the probability of decoding by mistake to another codeword from the same copy of the faded FC $S(n,\epsilon',a/\sigma)_{\mathbf{H}}$ or to a codeword in another copy. Hence, by using the union bound, we obtain the following:
\begin{align}
P_{e,ML}^{IC}(s_{\text{rc}}|\mathbf{H}) &\leq \left( P_{e,ML}^{FC}(s_{\text{rc}}|\mathbf{H}) + \sum_{i=1}^{n}{2Q\left(\frac{H_i\cdot b}{2\sigma}\right)} \right)\cdot 1_{\left\{H_{\min} > h_{\min}^{*}\right\}} + 1_{\left\{H_{\min} \leq h_{\min}^{*}\right\}}
\\ &\leq P_{e,ML}^{FC}(s_{\text{rc}}|\mathbf{H}) + 2nQ\left(\frac{h_{\min}^{*}\cdot b}{2\sigma}\right) + 1_{\left\{H_{\min} \leq h_{\min}^{*}\right\}}.
\end{align}
The average error probability over $S(n,\epsilon)_{\mathbf{H}}$ and $\mathbf{H}$ is then upper bounded by
\begin{equation}
\label{eq_IC_error_pro_ub_1}
P_{e,ML}^{IC} \leq P_{e,ML}^{FC} + 2nQ\left(\frac{h_{\min}^{*}\cdot b}{2\sigma}\right) + Pr\left\{H_{\min} \leq h_{\min}^{*}\right\}.
\end{equation}
Trivially we have
\begin{equation}
\label{eq_FC_error_pro_ub}
P_{e,ML}^{FC} \leq P_{e,DT}^{FC} \leq \epsilon',
\end{equation}
where $P_{e,DT}^{FC}$ is the average error probability of the FC using the suboptimal decoder on which the dependence testing bound is based.
\newline
By the union bound
\begin{equation}
\label{eq_error_pro_according_h_min_ub}
Pr\left\{H_{\min} \leq h_{\min}^{*}\right\} \leq nPr\left\{H \leq h_{\min}^{*}\right\}.
\end{equation}
Combining \eqref{eq_IC_error_pro_ub_1}, \eqref{eq_FC_error_pro_ub} and \eqref{eq_error_pro_according_h_min_ub} we get that
\begin{equation}
\label{eq_IC_error_pro_ub_2}
P_{e,ML}^{IC} \leq \epsilon' + 2nQ\left(\frac{h_{\min}^{*}\cdot b}{2\sigma}\right) + nPr\left\{H \leq h_{\min}^{*}\right\} \triangleq \epsilon.
\end{equation}

From \eqref{align_IC_NLD} and \eqref{eq_IC_error_pro_ub_2} we can see that for any large enough $n$, if we choose small enough $h_{\min}^{*}$, large enough $b$ relative to $h_{\min}^{*}/\sigma$ and large enough $a$ relative to $b$, then we will get an IC with average error probability which is upper bounded by $\epsilon$ and arbitrarily close to $\epsilon'$, and NLD which equals $\delta(n,\epsilon) \triangleq \delta^* - \sqrt{ \frac{V}{n} }Q^{-1}(\epsilon) + O\left(\frac{1}{n}\right)$.

Let us demonstrate this idea by an example. Suppose a regular fading distribution s.t. $f(h)\sim\frac{1}{h^{1-\alpha}}$ for small enough positive $h$ and for some $\alpha > 0$. Hence, $Pr\left\{H \leq h_{\min}^{*}\right\}=O\left(\left(h_{\min}^*\right)^{\alpha}\right)$. If we choose $h_{\min}^*(n) = \frac{1}{n^{\frac{2}{\alpha}}}, b(n) = \sigma\cdot n^{1+\frac{2}{\alpha}}$ and $a(n) = \sigma\cdot n^{2+\frac{2}{\alpha}}$, then we will get:
\begin{align}
\label{eq_IC_error_pro_ub_3}
\begin{aligned}
P_{e,ML}^{IC}
   &\leq \epsilon
\\ &\triangleq \epsilon' + 2nQ\left(\frac{h_{\min}^{*}(n)\cdot b(n)}{2\sigma}\right) + nPr\left\{H_{\min} \leq h_{\min}^{*}(n)\right\}
\\ &\leq \epsilon' + 2nQ\left(\frac{n}{2}\right) + O\left(\frac{1}{n}\right)
\\ &\leq \epsilon' + ne^{-\frac{n^2}{8}} + O\left(\frac{1}{n}\right)
\\ &= \epsilon' + O\left(\frac{1}{n}\right)
\end{aligned}
\end{align}
and
\begin{align}
\begin{aligned}
\delta(n,\epsilon,a(n)/\sigma,b(n))
   &= \delta(n,\epsilon',a(n)/\sigma) - \ln\left(1+\frac{b(n)}{a(n)}\right)
\\ &= \delta\left(n,\epsilon - O\left(1/n\right),a(n)/\sigma\right) + O\left(\frac{1}{n}\right)
\\ &= \delta^* - \sqrt{ \frac{V}{n} }Q^{-1}\left(\epsilon - O\left(1/n\right)\right)
+ O\left(\frac{1}{n} + \frac{1}{\sqrt{n}}\left(\frac{\sigma}{a(n)}\right)^{\frac{\alpha}{2}} + \left(\frac{\sigma}{a(n)}\right)^{\alpha}\right)
\\ &= \delta^* - \sqrt{ \frac{V}{n} }Q^{-1}(\epsilon) + O\left(\frac{1}{n}\right) \triangleq \delta(n,\epsilon).
\end{aligned}
\end{align}
Note that this operation can be done for any fixed $\epsilon > 0$ (or equivalently for any $\epsilon' > 0$).
\end{appendices}
\bibliographystyle{unsrt}

\end{document}